\documentclass[longauth]{aa} 

\usepackage{graphicx}
\usepackage{colortbl}
\usepackage{txfonts}

\begin{document} 

%-----------------------------------------------------------------------
% TITLE & AUTHORS
%-----------------------------------------------------------------------

   \title{Seeds of Life in Space (SOLIS).VII. Discovery of a cold dense methanol blob toward the L1521F VeLLO system\thanks{Based on observations carried out under project number L15AA with the IRAM-NOEMA interferometer. IRAM is supported by INSU/CNRS (France), MPG (Germany), and IGN (Spain).}}

\author{C. Favre 
          \inst{1}
          \and
          C. Vastel\inst{2}
          \and
          I. Jimenez-Serra\inst{3}
          \and
          D.  Qu\'enard\inst{4}
          \and
          P. Caselli\inst{5}
          \and
          C. Ceccarelli\inst{1}
          \and
          A. Chac\'on-Tanarro\inst{6}
          \and 
          F. Fontani\inst{7,5}
          \and 
          J. Holdship\inst{8,9}
          \and
          Y. Oya\inst{10}
          \and
          A. Punanova\inst{11}
          \and
          N. Sakai\inst{12}
          \and
          S. Spezzano\inst{5}
          \and
          S. Yamamoto\inst{10}
          \and
          R. Neri\inst{13}
          \and
          A. L\'opez-Sepulcre\inst{1,13}
          \and
          F. Alves\inst{5}          
          \and
          R. Bachiller\inst{6}          
          \and
          N. Balucani\inst{14,7,1}
          \and
          E. Bianchi\inst{1} 
          \and
          L. Bizzocchi\inst{5}         
          \and
          C. Codella\inst{7,1}
          \and
          E. Caux\inst{2}
          \and
          M. De Simone\inst{1}
          \and
          J. Enrique Romero\inst{1,15}
          \and
          F. Dulieu\inst{16}          
          \and
          S. Feng\inst{5}
          \and
          A. Jaber Al-Edhari\inst{1,17}
          \and
          B. Lefloch\inst{1}
          \and
          J. Ospina-Zamudio\inst{1}
          \and
          J. Pineda\inst{5}          
          \and
          L. Podio\inst{7}
          \and
          A. Rimola\inst{15}
          \and
          D. Segura-Cox\inst{5}
          \and
          I.~R. Sims\inst{18}
          \and
          V. Taquet\inst{7}
          \and
           L. Testi\inst{19,7}
          \and
          P. Theul\'e\inst{20}
          \and
          P. Ugliengo\inst{21}
          \and
          A.I. Vasyunin\inst{22,11}
          \and
          F. Vazart\inst{1}
          \and
          S. Viti\inst{8}
         \and
          A. Witzel\inst{1}
                                  }  

 \institute{
%1
Univ. Grenoble Alpes, CNRS, Institut de Plan\'etologie et d'Astrophysique de Grenoble (IPAG), 38000 Grenoble, France\\
\email{cecile.favre@univ-grenoble-alpes.fr}
\and
%2
IRAP, Universit\'e de Toulouse, CNRS, CNES, UPS, (Toulouse), France
\and
%3 
Centro de Astrobiolog\'ia (CSIC, INTA), Ctra. de Torrejon a Ajalvir km4, Torrej\'on de Ardoz, 28850 Madrid, Spain
\and
%4
School of Physics and Astronomy, Queen Mary University of London, 327 Mile End Road, London, E1 4NS, UK
\and
%5
Max-Planck-Institut f\"ur extraterrestrische Physik (MPE), 85748 Garching, Germany
\and
%6
Observatorio Astron\'omico Nacional (OAG-IGN), Alfonso XII 3, 28014, Madrid, Spain
\and
%7
INAF-Osservatorio Astrofisico di Arcetri, Largo E. Fermi 5, I-50125, Florence, Italy
\and
%8
Department of Physics and Astronomy, University College London, Gower Street, WC1E 6BT, London, UK  
\and
%9
Leiden Observatory, Leiden University, PO Box 9513, NL-2300 RA Leiden, The Netherlands
\and
%10
Department of Physics, The University of Tokyo, Bunkyo-ku, 113-0033 Tokyo, Japan
\and
%11
Ural Federal University, 620002, 19 Mira street, Yekaterinburg, Russia
\and
%12
RIKEN Cluster for Pioneering Research, 2-1, Hirosawa, Wako-shi, 351-0198 Saitama, Japan
\and
%13
Institut de Radioastronomie Millim\'etrique, 300 rue de la Piscine, Domaine
Universitaire de Grenoble, 38406, Saint-Martin d'H\`eres, France
\and
%14
Dipartimento di Chimica, Biologia e Biotecnologie, Via Elce di Sotto 8, 06123 Perugia, Italy
\and
%15
Departament de Qu\'{\i}mica, Universitat Aut\`onoma de Barcelona, 08193 Bellaterra, Catalonia, Spain
\and
%16
LERMA, Universit\'e de Cergy-Pontoise, Observatoire de Paris, 
PSL Research University, CNRS, Sorbonne Universit\'e, UPMC, Univ. Paris 06, 95000 Cergy Pontoise, France
\and
%17
University of AL-Muthanna, College of Science, Physics Department, AL-Muthanna, Iraq
\and
%18
Univ. Rennes, CNRS, IPR (Institut de Physique de Rennes) - UMR 6251, F-35000 Rennes, France
\and
%19
ESO, Karl Schwarzchild Srt. 2, 85478 Garching bei M\"unchen, Germany
\and
%20
Aix-Marseille Universit\'e, PIIM UMR-CNRS 7345, 13397 Marseille, France
\and
%21
Universit\`a degli Studi di Torino, Dipartimento Chimica Via Pietro Giuria 7, 10125 Torino, Italy
\and
%22
Engineering Research Institute {\it Ventspils International Radio Astronomy Centre} of Ventspils University of Applied Sciences, Inzenieru 101, Ventspils 3601, Latvia }

 %  \date{Received September 15, 1996; accepted March 16, 1997}

% \abstract{}{}{}{}{} 
% 5 {} token are mandatory

%------------------------------------------------------------------
%--------ABSTRACT--------------
%------------------------------------------------------------------
 
  \abstract
   {}
   {The SOLIS (Seeds Of Life In Space) IRAM/NOEMA Large Program aims at studying a set of crucial complex organic molecules in a sample of sources, with well-known physical structure, covering the various phases of Solar-type star formation. One representative object of the transition from the prestellar core to the protostar phases has been observed toward the Very Low Luminosity Object (VeLLO) called L1521F. This type of source is important to study to make the link between prestellar cores and Class 0 sources and also to constrain the chemical evolution during the process of star formation.}
   {Two frequency windows (81.6--82.6 GHz and 96.65--97.65 GHz) were used to observe the emission from several complex organics toward the L1521F VeLLO. These set-ups cover transitions of ketene (H$_2$CCO), propyne (CH$_3$CCH), formamide (NH$_2$CHO), methoxy (CH$_3$O), methanol (CH$_3$OH), dimethyl ether (CH$_3$OCH$_3$) and methyl formate (HCOOCH$_3$).}
   {Only 2 transitions of methanol (A$^+$, E$_2$) have been detected in the narrow window centered at 96.7 GHz (with an upper limit on E$_1$) in a very compact emission  blob ($\sim$ 7$^{\prime\prime}$ corresponding to $\sim$ 1000 au) toward the North-East of the L1521F protostar. The CS 2--1 transition is also detected within the WideX bandwidth. Consistently, with what has been found in prestellar cores, the methanol emission appears $\sim$1000 au away from the dust peak. The location of the methanol blob coincides with one of the filaments previously reported in the literature. The excitation temperature of the gas inferred from methanol is (10$\pm$2) K, while the H$_2$ gas density (estimated from the detected CS 2--1 emission and previous CS 5--4 ALMA observations) is a factor $>$25 higher than the density in the surrounding environment (n(H$_{2}$) $\geq$10$^{7}$ cm$^{-3}$).}
   {From its compactness, low excitation temperature and high gas density, we suggest that the methanol emission detected with NOEMA is {\it i)} either a cold and dense shock-induced blob, recently formed ($\leq$ few hundred years) by infalling gas or {\it ii)} a cold and dense fragment that may have just been formed as a result of the intense gas dynamics found within the L1521F VeLLO system.}
   \keywords{astrochemistry---line: identification---ISM: abundances---ISM: molecules---ISM: individual objects (L1521F)}

   \maketitle
%
%-------------------------------------------------------------------

%===============================================================
%
%-----------------------------------------------------------------------------------------------------------------------------
%--------INTRODUCTION --------------------------------------------
%------------------------------------------------------------------
\section{Introduction}
\label{sec:introduction}
%------------------------------------------------------------------
About a decade ago, the {\it Herschel} satellite revealed that the ISM is organised in a complex network of filamentary structures or filaments \citep{Andre:2010}. These filaments are believed to undergo gravitational fragmentation into multiple fragments that subsequently form dense and cold prestellar cores \citep[see e.g][]{Andre:2019}. 

It is well established that low-mass protostars are born within prestellar cores. However, the transition between a prestellar core and a protostar (called first hydrostatic core phase or FHSC) is poorly known. The stage of FHSC starts when the density of the central object increases enough (via accretion) to become opaque to radiation, and lasts until its temperature reaches 2000 K forcing the dissociation of H$_2$ \citep{Larson:1969,Masunaga:1998,Masunaga:2000}. Because of their short life times \citep[0.5-50 kyr,][]{Omukai:2007,Tomida:2010,Commercon:2012}, the identification of FHSCs is challenging.   

Several observational studies have attempted to search and identify FHSCs \citep{Dunham:2008,Dunham:2011,Chen:2010,Chen:2012,Enoch:2010,Schnee:2012,Murillo:2013}. Some of them have proposed that Very Low Luminosity Objects (or VeLLOs, with L$_{bol}$$\leq$0.1 L$_\odot$) could be FHSCs \citep[as e.g.][]{Enoch:2010}. However, their true nature is still under debate. For example, \citet{Vorobyov:2017} argue that the majority of the VeLLOs should be in the evolved Class I protostars phase, where protostars have already grown in mass via "cold accretion" \citep[i.e. a phenomenon by which the accreting gas provides a very low entropy to the protostar;][]{Hosokawa:2011}.
 Interestingly \citet{Tokuda:2017} have found that the very low luminosity protostar in the L1521F system has a central stellar mass of $\sim$~0.2M$_\odot$ and have suggested that this finding can likely be explained by the cold accretion model.
Therefore, studies of VeLLOs not only are important to understand the earliest evolutionary stages in the formation process of low-mass stars, but also they may represent the missing link between the prestellar core phase and the Class 0 phase.

L1521F \citep[also known as MC27, see][]{codella1997,mizuno1994,onishi1996,onishi1998,onishi1999,onishi2002} is one of the densest cores in the nearby \citep[$\sim$136 pc:][]{Maheswar:2011} Taurus molecular cloud. It was originally classified as a starless core and was the subject of many studies, sharing many similarities with the prototypical prestellar core L1544. \citet{crapsi2004} observed L1521F in dust emission at 1.2 mm and in two transitions each of N$_2$H$^+$, N$_2$D$^+$, C$^{18}$O and C$^{17}$O. They measured a molecular hydrogen number density n(H$_2$) $\sim$ 10$^6$ cm$^{-3}$ and a CO depletion factor, integrated along the line of sight, of f$_D$ = $\rm 9.5 \times 10^{-5}/x_{obs}(CO) \sim 15$, similar to that derived toward the prestellar core L1544. The N(N$_2$D$^+$)/N(N$_2$H$^+$) column density ratio is $\sim$ 0.1, a factor of about 2 lower than that found in L1544. The N$_2$H$^+$ and N$_2$D$^+$ linewidths in the core center are $\sim$ 0.3 km~s$^{-1}$, significantly larger than in other more quiescent Taurus starless cores but similar to those observed toward the center of L1544. From all this, \citet{crapsi2004,crapsi2005} concluded that L1521F is less evolved than L1544, but, in analogy with the latter core, it is approaching the "critical" state. 

The view on the physical nature of the L1521F core changed thanks to the high sensitivity of the {\it Spitzer} telescope, which detected a very low luminosity protostar (<0.07 L$_{\odot}$) in a very dense region \citep[10$^6$ cm$^{-3}$;][]{bourke2006}. L1521F-IRS is currently classified as a VeLLO \citep[see e.g.,][]{Young:2004,Dunham:2006,Lee:2009a}.

Subsequent interferometric observations carried out with the SMA in $^{12}$CO (2--1) and 1.3 mm continuum emission, spatially resolved a compact but poorly collimated molecular outflow associated with L1521F-IRS \citep{takahashi2013}. This suggests that L1521F is at the earliest protostellar stage (<10$^4$ yr). In addition, higher-angular resolution observations carried out with the IRAM-PdBI and ALMA showed that this source in fact splits into a small cluster of cores, MMS-1, MMS-2 and MMS-3 \citep[see][]{Maury:2010,tokuda2014}, where MMS-1 coincides with the location of the L1521F-IRS {\it Spitzer} source.  The SMA observations of \citet{takahashi2013} also unveiled another object in the region with evidence of compact CO blueshifted and redshifted components toward the northeast of L1521F-IRS, called L1521F-NE. However, no driving source has been detected in either millimeter continuum emission with PdBI/SMA or infrared emission with {\it Spitzer}, as confirmed by ALMA Cycle 1 observations \citep[][]{tokuda2014,Tokuda:2016}. From this, \citet{takahashi2013} derived a mass detection limit of 10$^{-4}$ M$_{\odot}$ for L1521F-NE.

In this paper, we present new interferometric observations of L1521F carried out with NOEMA (Northern Extended Millimetre Array) to investigate the molecular complexity of its identified VeLLO. This work is part of the NOEMA large program SOLIS \citep[Seeds of Life in Space:][]{Ceccarelli:2017}, aimed at studying the formation of complex organic molecules across all stages of star formation \citep{Ceccarelli:2017}. In Section 2, the details of the observations, the data reduction procedure and Gaussian fitting of the spectra are presented. Section 3 presents the results of the Gaussian fitting, velocity gradients, rotational temperatures, and column density calculations of the CH$_3$OH and CS molecular lines detected toward L1521F with NOEMA. In Section 4, we discuss the results and possible origins of the methanol-rich blob located $\sim$1000 au away from the L1521F source. 

%===============================================================
%
%-----------------------------------------------------------------------------------------------------------------------------
%-------------------------------------------- OBSERVATIONS -------------------------------------------------
%-----------------------------------------------------------------------------------------------------------------------------
\section{Observations and data reduction}
\label{sec:observations}
%-----------------------------------------------------------------------------------------------------------------------------
%
\subsection{IRAM Observations}

\subsubsection{NOEMA Observations}

The IRAM-NOEMA observations were carried out in C and D configurations between September 2016 and January 2017 under average weather conditions (pwv = 1--10 mm) toward L1521F (${\rm \alpha_{2000} = 04^h28^m38.99^s, \delta_{2000} = 26\degr51\arcmin35.6\arcsec}$). The rest frequencies have been shifted with respect to the V$_{LSR}$ of the source ($\sim$ 6.4--6.6 km~s$^{-1}$). The primary beam size was 52${\rm \arcsec}$, the synthesized beam was 2.72$\arcsec$ $\times$  2.37$\arcsec$ at a position angle 29$\degr$. The data were obtained with the narrowband correlator with a spectral resolution of 39 kHz, corresponding to a velocity resolution of 0.12 km~s$^{-1}$. The system temperatures were 
50--110~K. 
The nearby sources 3C454.3 and J0438+300 were used as bandpass and gain (phase and amplitude) calibrators, respectively. The absolute flux calibration was performed via observation of the quasar MWC34 (1.03~Jy).\\
Two methanol transitions have been detected in the narrowband correlator: E$_2$ ${\rm 2_{1,2}-1_{1,1}}$ (96.739362 GHz), A$^+$ ${\rm 2_{0,2}-1_{0,1}}$ (96.741375 GHz)  in the North-East position of L1521F while E$_1$ ${\rm 2_{0,2}-1_{0,1}}$ (96.744550 GHz) is marginally detected. The emission is clearly extended (see Section~3.3).
Along with the E$_2$ ${\rm 2_{1,2}-1_{1,1}}$ (96.739362 GHz), A$^+$ ${\rm 2_{0,2}-1_{0,1}}$ (96.741375 GHz), and E$_1$ ${\rm 2_{0,2}-1_{0,1}}$ (96.744550 GHz) methanol transitions, the dimethyl ether (E and A CH$_3$OCH$_3$ ${\rm 5_{5,1}-4_{4,0}}$ at 95.85 GHz) and methyl formate (E-CH$_3$OCHO ${\rm 5_{4,1}-5_{3,3}}$ at 96.94 GHz and A-CH$_3$OCHO ${\rm 17_{5,12}-17_{4,13}}$ at 97.20 GHz) lines were observed within the same spectral setup with the narrowband correlator. Nonetheless, these molecular species \citep[as well as the other targeted COMs, see][]{Ceccarelli:2017} are not detected in the map at high spectral resolution (rms $\sim$ 3.8 mJy~beam$^{-1}$). 
In addition, the spectral range of WideX was 95.85--99.45 GHz and the CS(2--1) line at 97.98 GHz line was detected with a spectral resolution of 1950 kHz (6.0 km~s$^{-1}$) while the SO(2$_3$--1$_2$) line at 99.30 GHz was not detected with a rms $\sim$0.3-0.4 mJy~beam$^{-1}$ and a beam size of about 3$\arcsec$ $\times$  2.6$\arcsec$ (PA: 24$\degr$).

In addition, observations at about 82~GHz in C and D configurations were also performed toward L1521F between September and November 2016 with 8 antennas. However, only the continuum emission is detected in these data set (see Section 3.1). None of the targeted lines \citep[see][for further details]{Ceccarelli:2017} have been detected.

\subsubsection{IRAM-30m Observations}

In this paper, we also make use of IRAM-30m observations for recovering the most extended emission. The single-dish observations were carried out in 2016 August under good weather conditions (pwv of about 1-2 mm). The on-the-fly maps were obtained with the EMIR 090 (3 mm band) heterodyne receiver in position switching mode, using the FTS backend with a spectral resolution of 50 kHz; this corresponds to a velocity resolution of 0.15 km~s$^{-1}$ at the frequency of 96.74 GHz. The angular resolution was 25.6${\rm \arcsec}$. The ${\rm 3\arcmin \times 3\arcmin}$ maps were centered at the dust emission peak (${\rm \alpha_{2000} = 04^h28^m39.8^s, \delta_{2000} = 26\degr51\arcmin35\arcsec}$). The pointing accuracy of the 30m antenna was better than 1${\rm \arcsec}$. The system temperature was 157 K. A detailed description of the data will be given in an upcoming paper (Spezzano et al. in prep). 

\subsection{Data reduction}

The calibration, imaging and cleaning of the NOEMA data were performed using the CLIC and MAPPING packages of the {\sc GILDAS}\footnote{https://www.iram.fr/IRAMFR/GILDAS/} software (July 2018 version). We note that the images have been corrected for primary beam attenuation. Regarding the single dish data, the reduction has been performed with the {\sc GILDAS-CLASS} package.

%+++++++++++++++++++++
%++++ FIGURE 1 +++++
 \begin{figure}
   \centering
   \includegraphics[width=1.0\hsize]{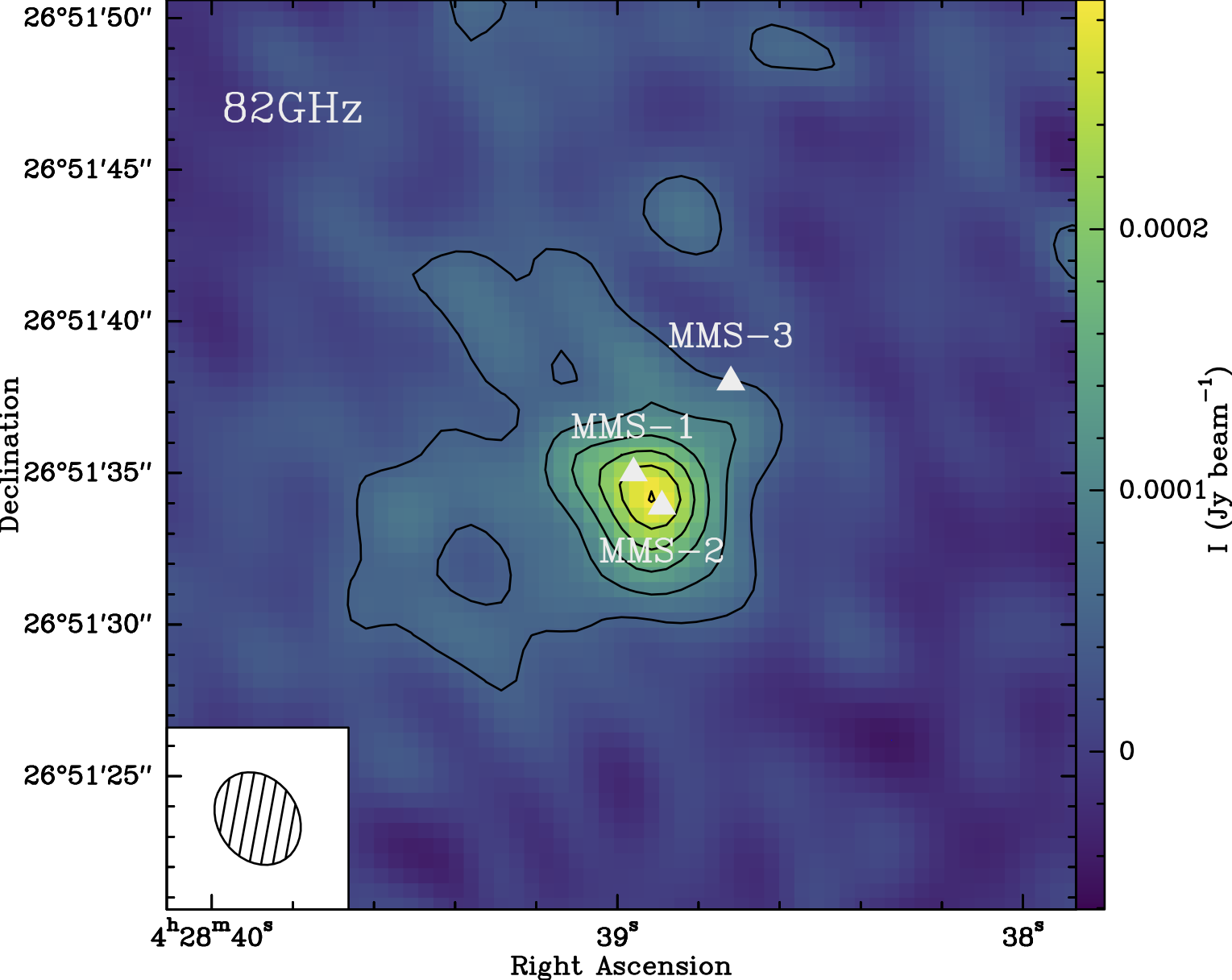}
   \includegraphics[width=1.0\hsize]{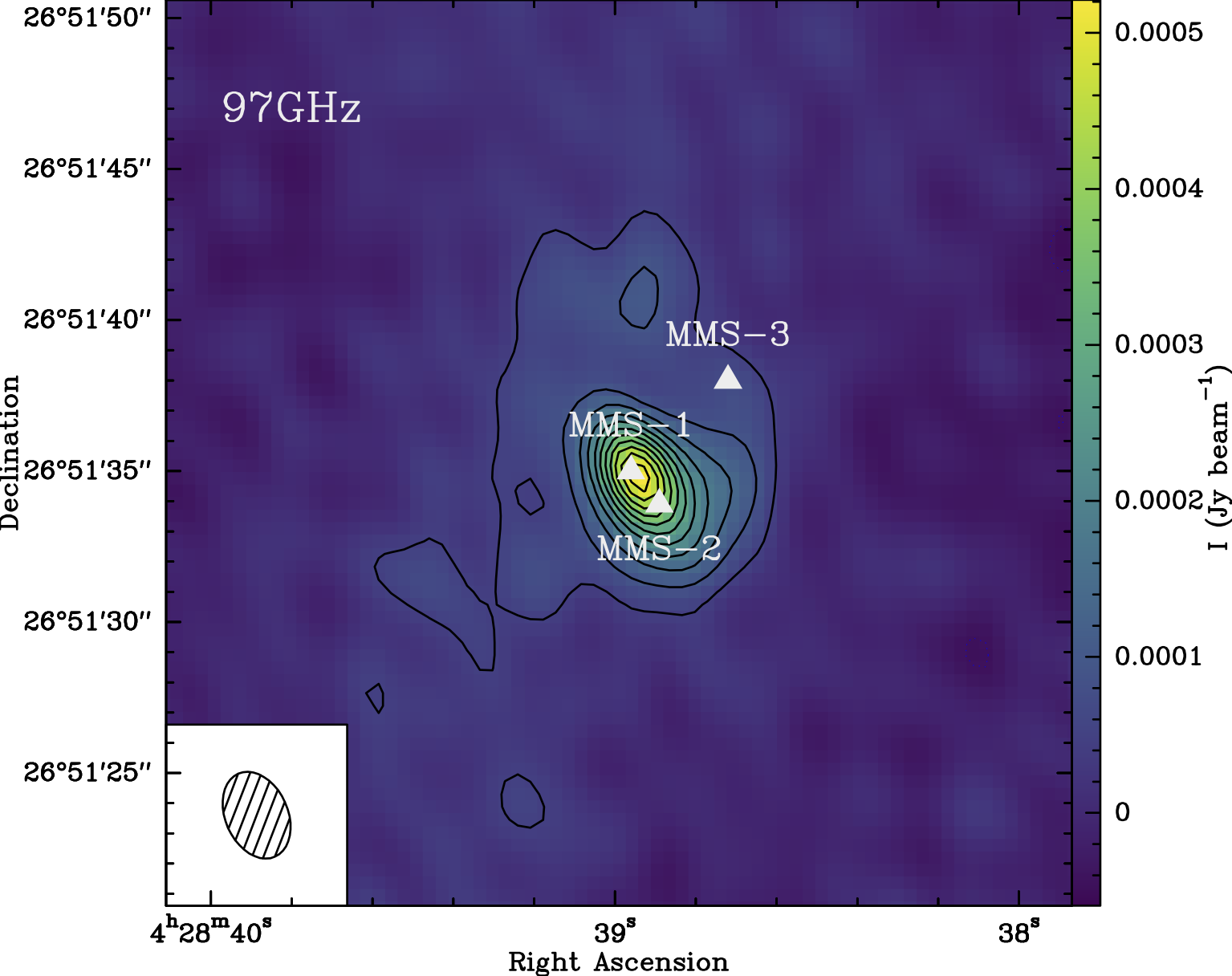}
   \caption{82 GHz (top panel: 3.65 mm) and 97 GHz (bottom panel: 3.1 mm) continuum emission as observed with NOEMA toward L1521F. The first contour and the level step are at 3$\sigma$ (where 1$\sigma$ = 1.6$\times$10$^{-5}$ and 1.5$\times$10$^{-5}$ Jy~beam~$^{-1}$ at 82 and 97 GHz, respectively). The white triangles indicate the positions of the MMS-1, MMS-2 and MMS-3 sources (see Section~1). Synthesized beams are shown in the bottom-left corner of the panels.}
   \label{fg1}
 \end{figure}
%+++++++++++++++++++++

\subsection{Missing flux and Data merging}

To estimate the portion of flux that is missing by the interferometer (due to spatial filtering), we compared the NOEMA and IRAM-30m data. In that context, the NOEMA data were convolved with a Gaussian beam similar to that of the 30m-data (i.e. $\sim$26$\arcsec$ at 96~GHz) and then smoothed  to the same spectral resolution as that of the 30m observations. By comparing the peak intensities in direction of the methanol emission peak, we estimate that more than 80$\%$ of the methanol emission is resolved out. 

To recover the missing flux, we merged the 30m with the NOEMA data via the use of a routine in the {\sc GILDAS-MAPPING} package. The resulting data cubes have a velocity resolution of 0.12 km~s$^{-1}$.The rms of the resulting spectral data cubes varies from 4  to 15 mJy/beam. The synthesized beam of the combined data cube is 2.8$^{\prime\prime}$ $\times$ 2.4$^{\prime\prime}$ at a position angle of 29$^{\degr}$, with a pixel size of 0.53$^{\prime\prime}$ $\times$ 0.53$^{\prime\prime}$.

%=================================================
%
%-----------------------------------------------------------------
%-------------------------------------------- RESULTS-------------
%-----------------------------------------------------------------
\section{Spatial distribution}
\label{sec:results}
%------------------------------------------------------------------

\subsection{Continuum emission}

\begin{figure}[h!]
   \includegraphics[width=1\hsize]{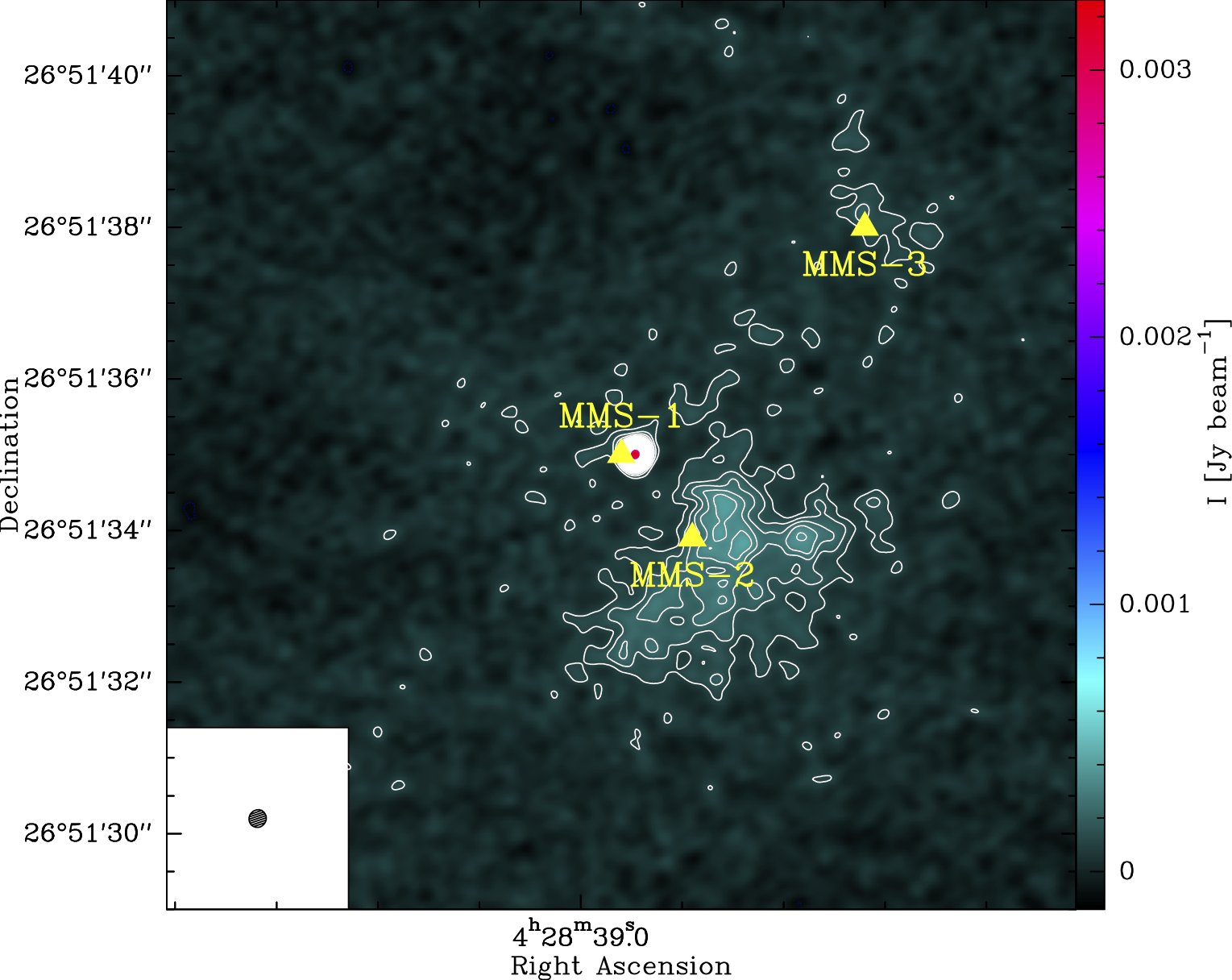}
  \caption{0.87 mm continuum map observed with ALMA (ADS/JAO.ALMA$\#$2015.1.00340.S, PI K. Tokuda). The first contour is at 3$\sigma$ and the level step at 2$\sigma$ (where 1$\sigma$ = 3.3$\times$10$^{-5}$ Jy~beam~$^{-1}$). The synthesized beam is shown in the bottom-left corner. The positions of sources MMS-1, MMS-2, MMS-3 reported by \citet{tokuda2014} are indicated as yellow triangles.}
   \label{fg2}
 \end{figure}

We present in Fig. \ref{fg1} the continuum maps at 3.1 mm (97 GHz) and 3.6 mm (82 GHz) along with the location of the MMS-1, MMS-2 and MMS-3 sources. \\
Surprisingly, MMS-3 is not detected in our maps, although it is barely detected at the 3$\sigma$ level at 0.87 mm and 1.2 mm using ALMA observations \citep{tokuda2014,Tokuda:2016}  and, detected at the 5$\sigma$ level,  as shown in Figure~\ref{fg2}, with ALMA Cycle 3 observations\footnote{The ALMA Cycle 3 continuum data were kindly given to us by K. Tokuda.} carried out at 0.87 mm (project code: ADS/JAO.ALMA$\#$2015.1.00340.S, PI K. Tokuda. For further details on the data reduction, see \citealt{Tokuda:2017,Tokuda:2018}).

%+++++++++++++++++++++
%++++ FIGURE 2 +++++
%+++++++++++++++++++++

Regarding the MMS-1 and MMS-2 sources, the present resolution of our NOEMA observations does not allow us to disentangle between the positions MMS-1 and MMS-2 (see Fig.~\ref{fg1}).  The total (MMS-1 $+$ MMS-2) measured flux density per synthesized beam, S$_{\nu}$, is of about 0.03 mJy~beam$^{-1}$ at 82~GHz and 0.05~mJy~beam$^{-1}$ at 97~GHz. 
Finally, it is interesting to note that MMS-2 is also detected at 230~GHz in the CALYPSO IRAM-PdBI survey by \citet{Maury:2019} but not in the CALYPSO pilot program performed at the same frequency \citep{Maury:2010}. We infer that the pilot data along with their calibration and reduction were preliminary.

\subsection{Methanol channel emission maps}

In this sub-section, we will only present  the resulting line emission obtained through the combined NOEMA and IRAM-30m data for the two following detected methanol transitions: E$_2$ ${\rm 2_{1,2}-1_{1,1}}$ at 96.739362 GHz and A$^+$ ${\rm 2_{0,2}-1_{0,1}}$ at 96.741375 GHz. We note that the higher energy level E$_1$ ${\rm 2_{0,2}-1_{0,1}}$ line at 96.744550 GHz line (E$_{up}$ = 20.1 K) is marginally detected (with an rms level of 3.6~mJy~beam$^{-1}$ or 0.08~mK). 

\begin{figure}
   \centering
   \includegraphics[width=1\hsize,clip=true,trim=0 0 0 0]{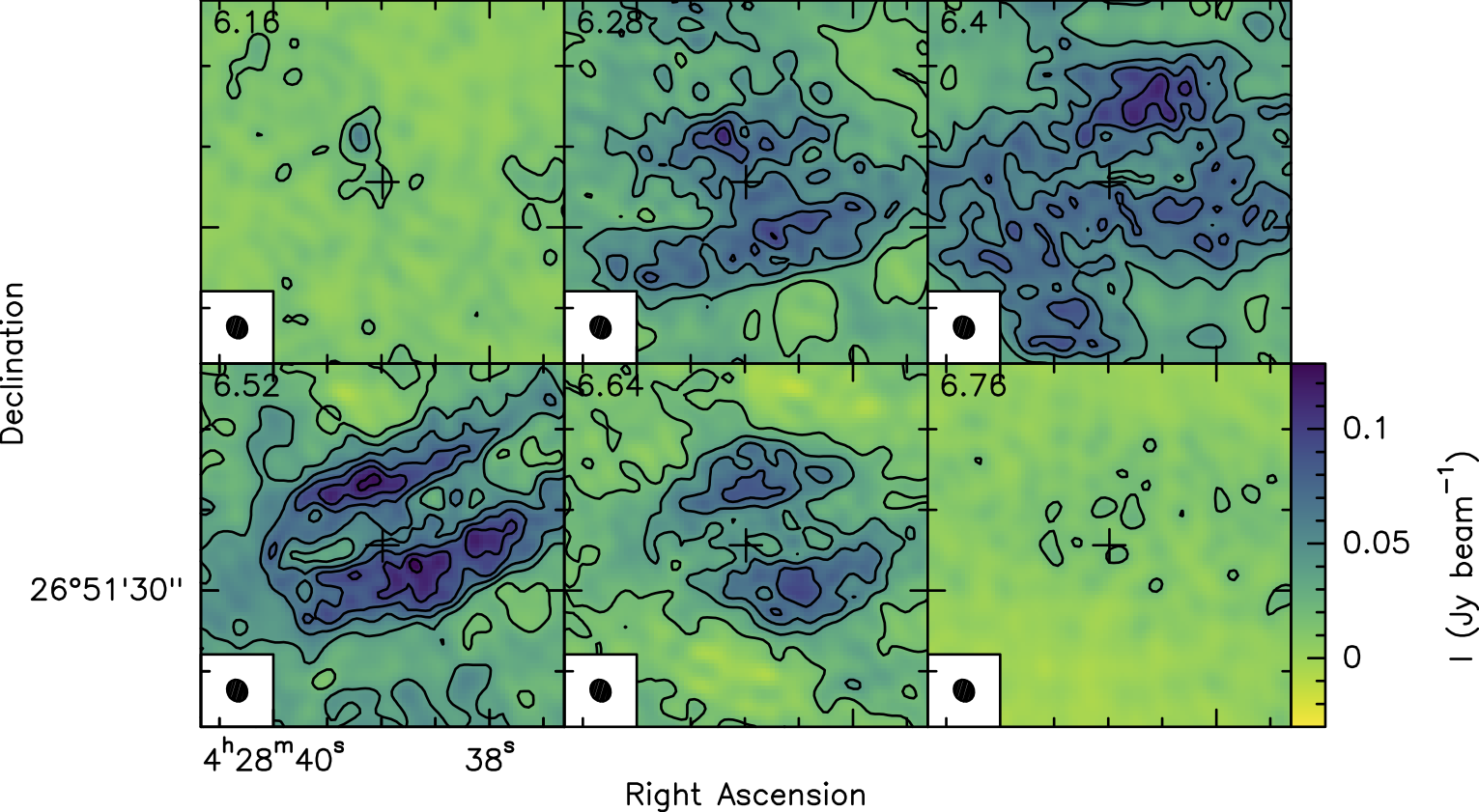}
   \includegraphics[width=1\hsize,clip=true,trim=0 0 0 0]{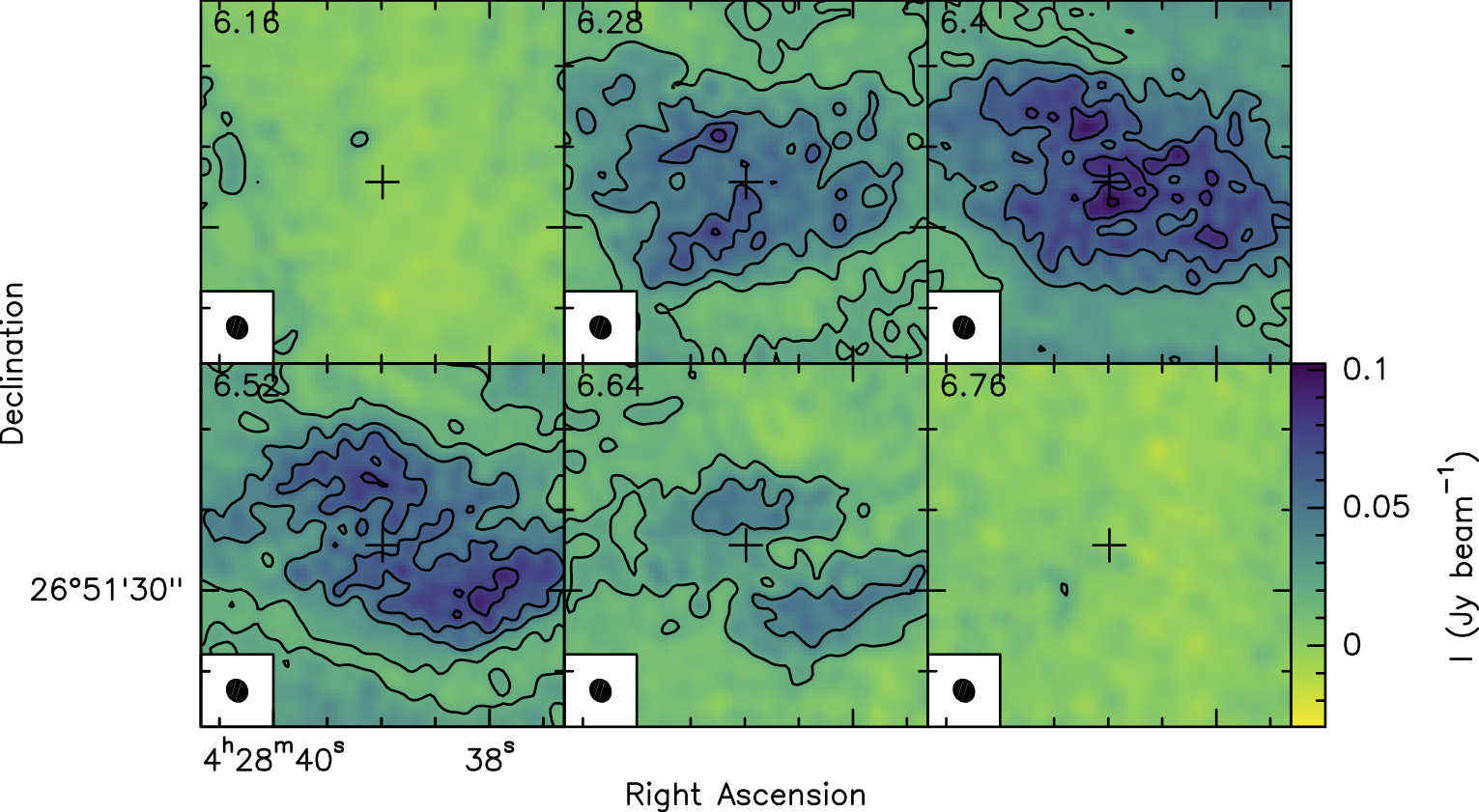}
   \caption{Velocity-channel maps of the A$^+$ (top) and E$_2$-CH$_3$OH (bottom) emission. The lowest contour starts at 20 mJy/beam, with a step of 20 mJy/beam.}
   \label{fg3}
 \end{figure}

Figure \ref{fg3} shows the channel emission maps for the A$^+$-- and E$_2$--CH$_3$OH lines. The respective emission presents both an arc-like structure.  A similar filamentary/arc-like structure have been previously observed in this source at  the same scale by \citet{tokuda2014} for HCO$^{+}$  (J$=$3--2).
In addition, at about 5-6 km~s$^{-1}$ the $^{12}$CO (J$=$3--2) emission is tracing an arc-like filamentary structure around L1521F \citep{Tokuda:2016} which is similar to that seen in HCO$^{+}$ and methanol \citep[see Fig.~6 from][and Fig.~\ref{fg3}]{Tokuda:2016}. It is interesting to note that CH$_3$OH ring-like distribution is also seen in other prestellar cores, such as TUKH122 \citep[which is on the verge of star formation, see][]{Ohashi:2018}.
In that light, \citet{tafalla2004} suggest that the ring-like morphology for CH$_3$OH in prestellar cores is due to depletion of C-bearing species close to the dust emission peak. 
Interestingly enough, \citet{Punanova:2018} has observed a centrally peaked  emission fragment for CH$_3$OH around the L1544 prestellar core center, and inferred that the methanol emission could arise from an accretion shock. 
Such structures are likely the result of dynamical gas interaction such as fragmentation \citep[see][]{tokuda2014}. In that context, we note that similar structures have been reproduced  by hydrodynamical simulations with and without magnetic field \citep{Matsumoto:2015,Matsumoto:2017}. Turbulence, injected by protostellar feedback, may indeed play a crucial role during fragmentation, different from what can be found in massive disks \citep{Larson:1987,Machida:2008}.

%+++++++++++++++++++++
%++++ FIGURE 3 +++++
%+++++++++++++++++++++

%+++++++++++++++++++++
%++++ FIGURE 4 +++++
\begin{figure*}
   \centering
   \includegraphics[width=0.49\hsize]{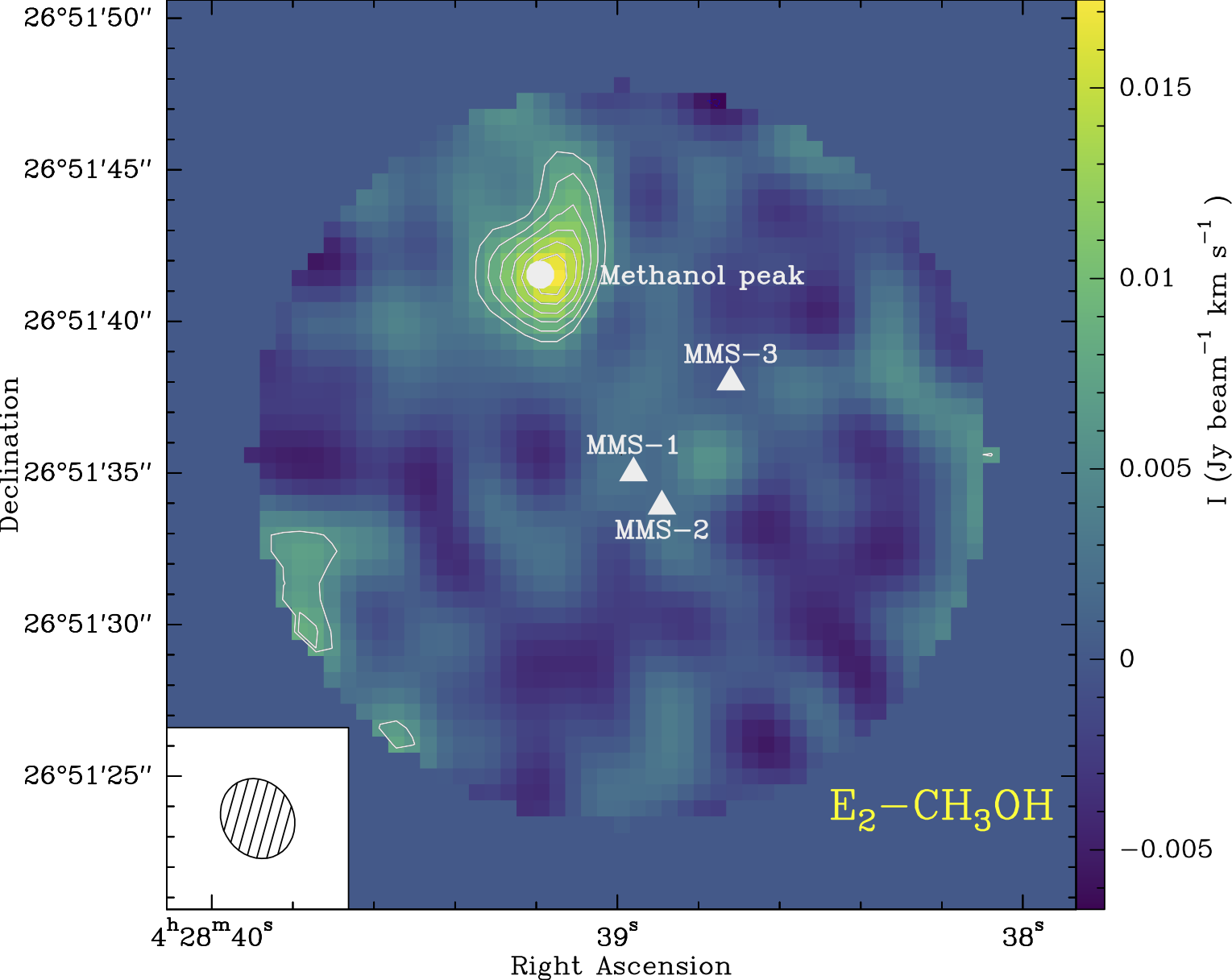}
   \includegraphics[width=0.49\hsize]{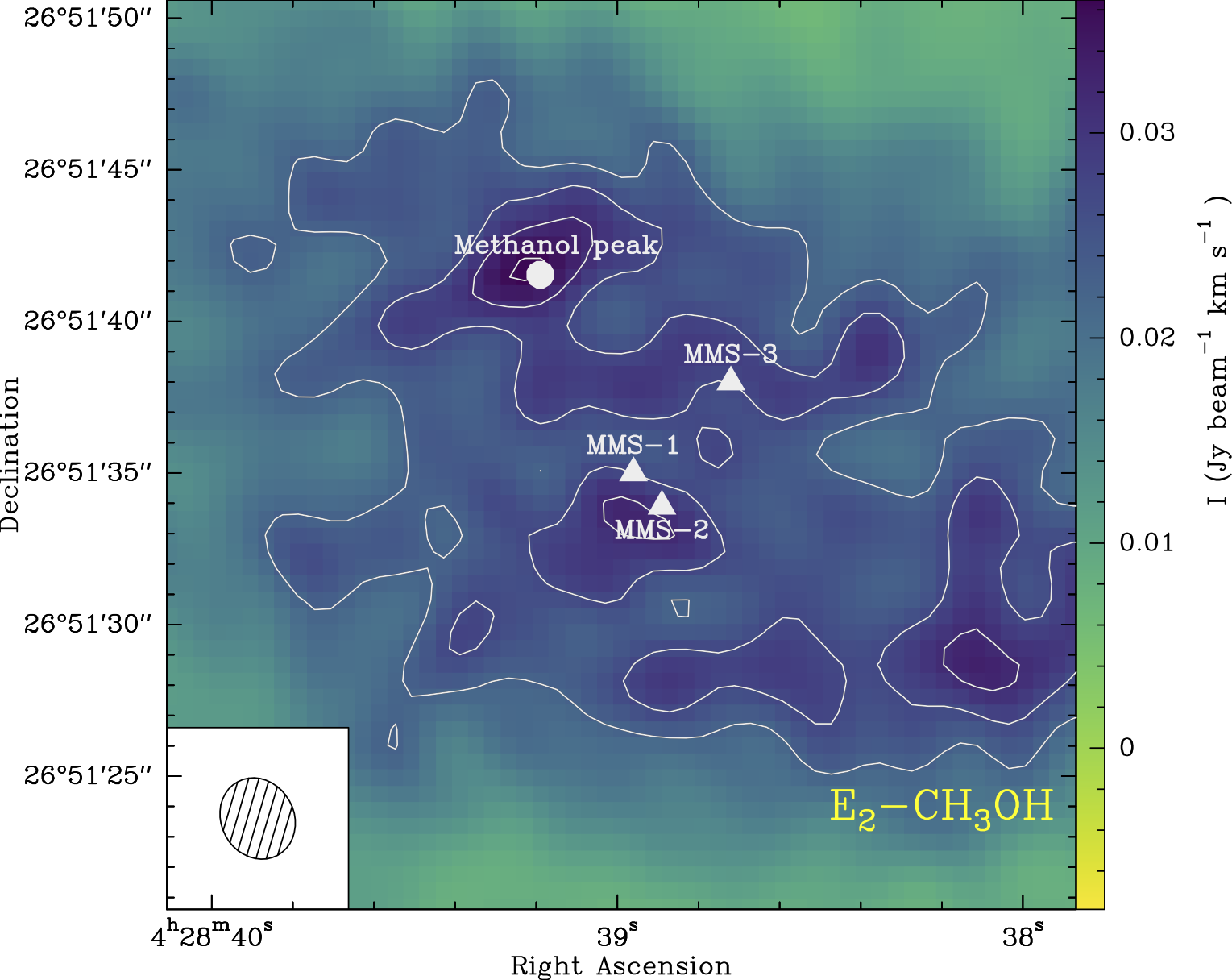}
   \includegraphics[width=0.49\hsize]{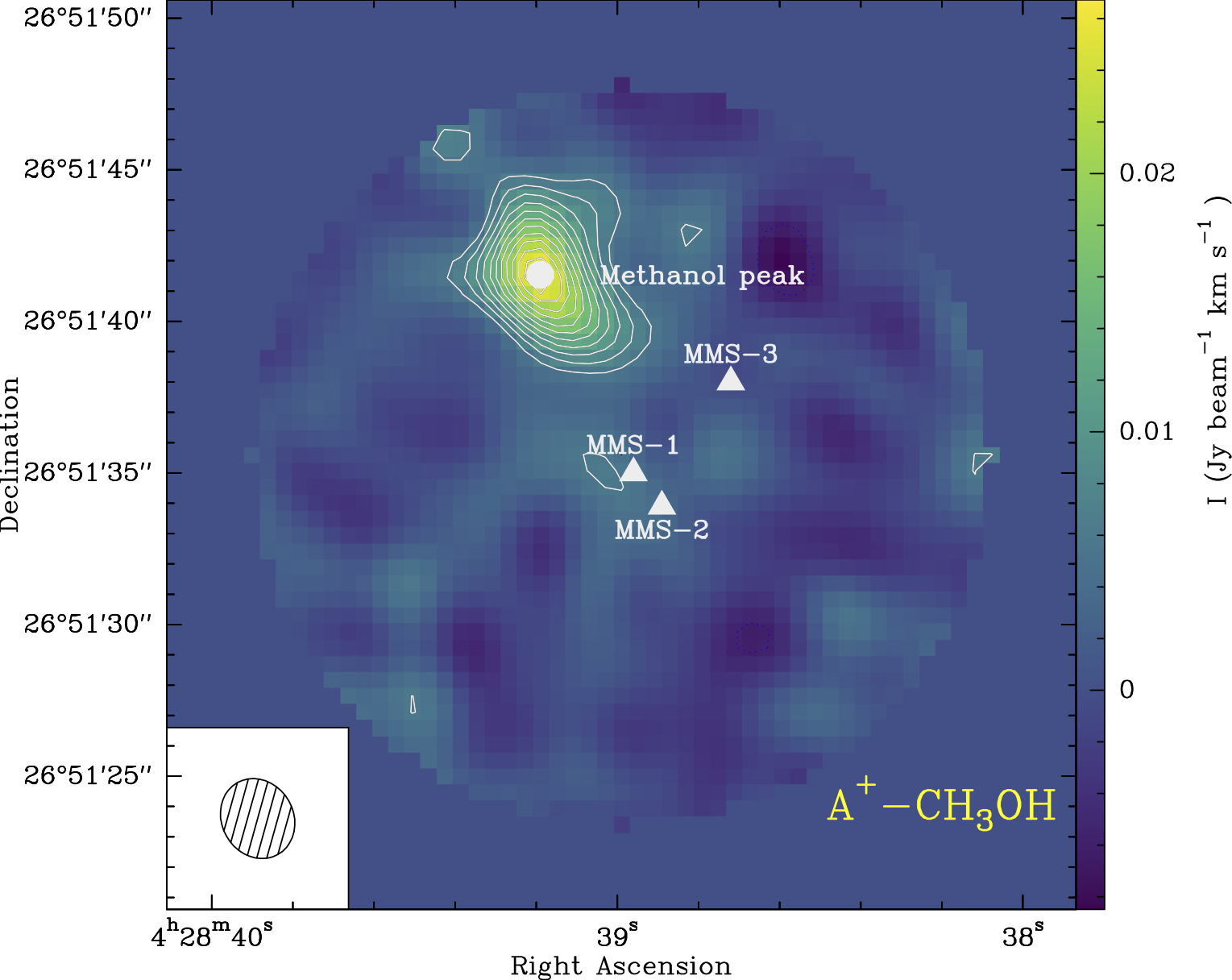}
   \includegraphics[width=0.49\hsize]{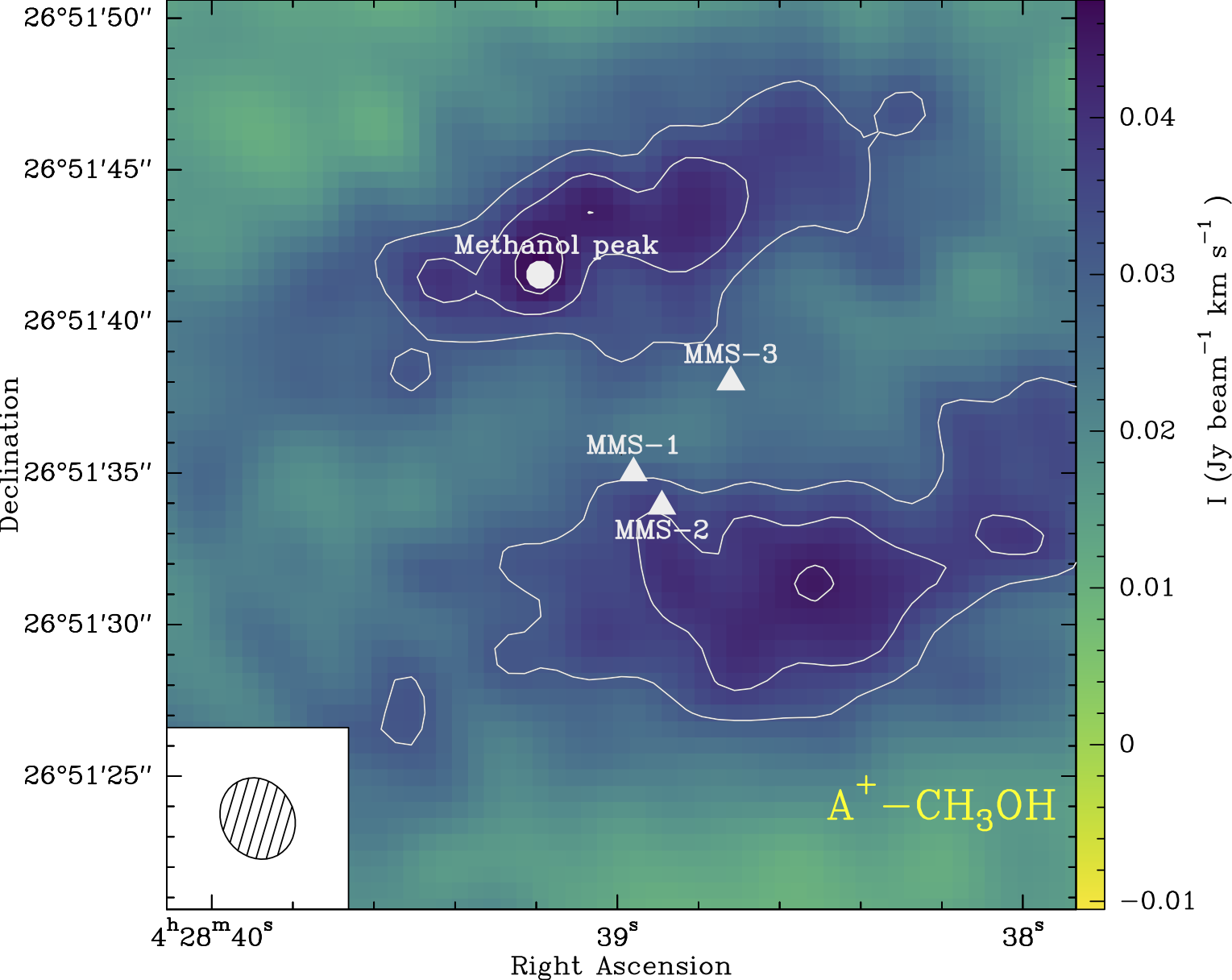}
   \includegraphics[width=0.49\hsize]{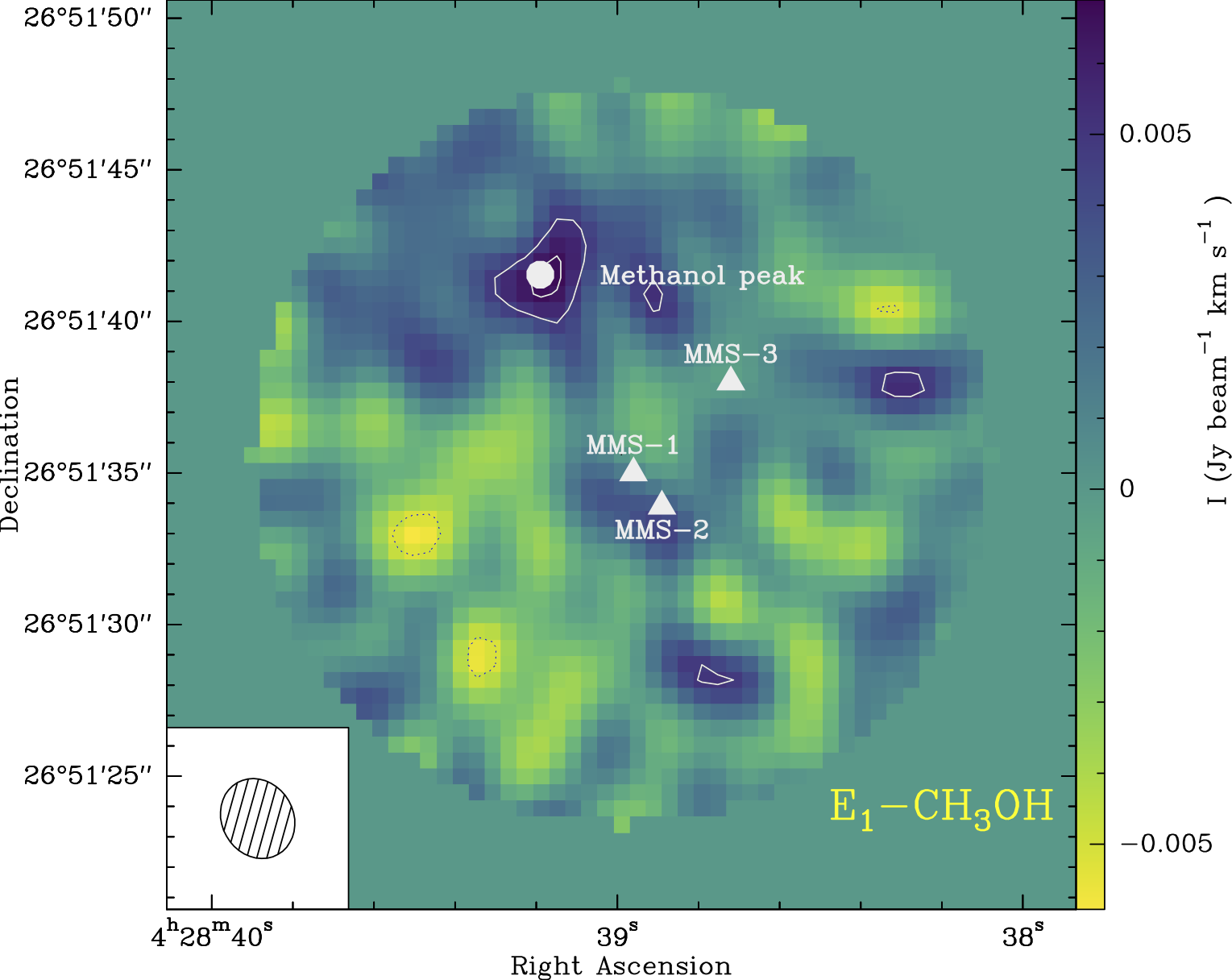}  
   \includegraphics[width=0.49\hsize]{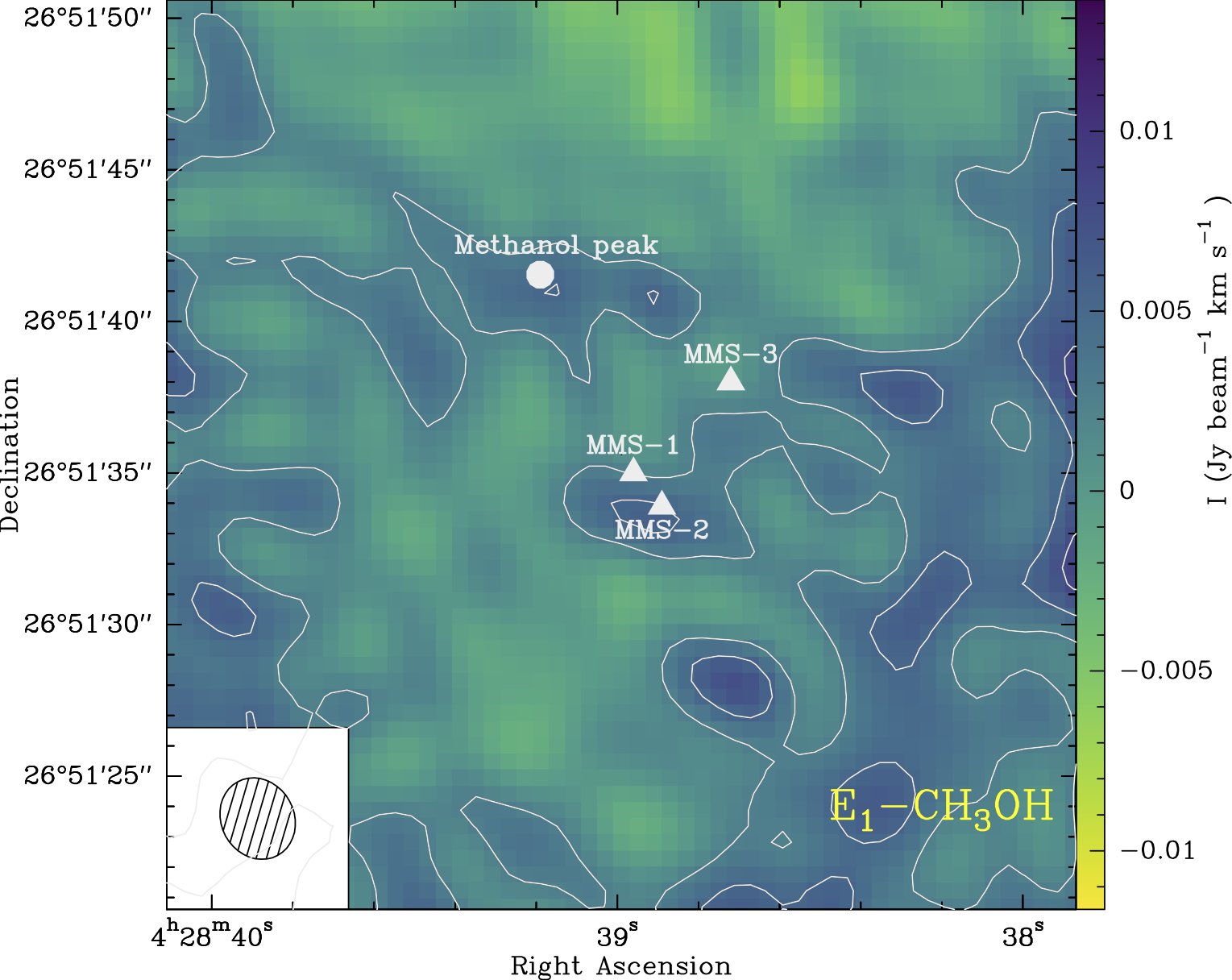}
   \caption{Methanol integrated emission maps (corrected from the primary beam attenuation). {\it Top left}: E$_2$-CH$_3$OH integrated intensity emission map over the line profile. The first contour is at 3$\sigma$ and the level step at 1$\sigma$ (where 1$\sigma$ = 1.9 mJy~beam~$^{-1}$~km~s$^{-1}$). {\it Top right}: CH$_3$OH-E  moment 0 maps from the combined IRAM-30m and NOEMA data. The first contour is at 5$\sigma$ and the level step at 1$\sigma$ (where 1$\sigma$ = 4.5 mJy~beam~$^{-1}$~km~s$^{-1}$). {\it Middle left}: A$^+$--CH$_3$OH integrated intensity emission map over the line profile. The first contour is at 3$\sigma$ and the level step at 1$\sigma$ (where 1$\sigma$ = 1.9 mJy~beam~$^{-1}$~km~s$^{-1}$). {\it Middle right}: A$^+$--CH$_3$OH  moment 0 maps from the combined IRAM-30m and NOEMA data. The first contour is at 5$\sigma$ and the level step at 1$\sigma$ (where 1$\sigma$ = 6.3 mJy~beam~$^{-1}$~km~s$^{-1}$). {\it Bottom left}: E$_1$--CH$_3$OH integrated intensity emission map over the line profile. The first contour is at 3$\sigma$ and the level step at 1$\sigma$ (where 1$\sigma$ = 1.6 mJy~beam~$^{-1}$~km~s$^{-1}$). {\it Bottom right}: E$_1$--CH$_3$OH  moment 0 maps from the combined IRAM-30m and NOEMA data. The first contour is at 1$\sigma$ (2.6 mJy~beam~$^{-1}$~km~s$^{-1}$).}
   \label{fg4}
 \end{figure*}
 %+++++++++++++++++++++

\subsection{Distribution of the methanol emission}

We present in Fig. \ref{fg4} the integrated intensities of the A$^+$, E$_1$ and E$_2$ methanol lines (left panel) observed with NOEMA only. The CH$_3$OH emission is compact and it peaks toward the North-East position of L1521F at coordinates ${\rm \alpha_{2000} = 04^h28^m39.164^s, \delta_{2000} = 26\degr51\arcmin41.49\arcsec}$). Surprisingly, this position is not associated with any of the 3 MMS sources. The source size is about 7$^{\prime\prime}$, corresponding to a 950 au size at a distance of 136 pc ($\sim$ 5~10$^{-3}$ pc).

In Fig. \ref{fg4} (right panels) we present the combined 30m-NOEMA images of the CH$_3$OH lines (see Section 2.3). The maps show that methanol is indeed extended and distributed in a ring-like structure around the Spitzer continuum source MMS-1, with V$_{LSR}$=6.4 km~s$^{-1}$. The emission is still brightest at the location of the CH$_3$OH peak that appears in the NOEMA-only images. This methanol peak (hereafter called {\it methanol blob}) resembles the methanol emission peak found toward the L1544 prestellar core, which is also located in the Taurus molecular cloud \citep[hence at the same distance as L1521F; see][]{bizzocchi2014}. Incidentally, we note that in the Taurus Molecular Cloud-1, the CH$_3$OH peak is also shifted from the denser part \citep{Soma:2015}.

Interestingly enough, \citet{Tokuda:2018} have recently focused on the large scale morphology and kinematics of the molecular gas around the protostar to understand the dynamical nature of the system, using $^{12}$CO (3--2), $^{12}$CO (2--1) and C$^{18}$O (2--1). From their analysis, MMS-2 is located south-west of the protostar in a warm filament with a 60~K kinetic temperature at a velocity range of 4.45--5.30 km~s$^{-1}$. The emission probed with NOEMA seems to be located at the intersection between 3 thin, cold (10--30 K) and dense filaments (n $\sim$ 10$^6$ cm$^{-3}$). We cannot exclude that the observed methanol blob could be part of the filamentary structure seen in CO and could be the result of accreting material.

%+++++++++++++++++++++
%++++ FIGURE 5 +++++
 \begin{figure*}[h!]
   \centering
 \includegraphics[width=0.45\hsize]{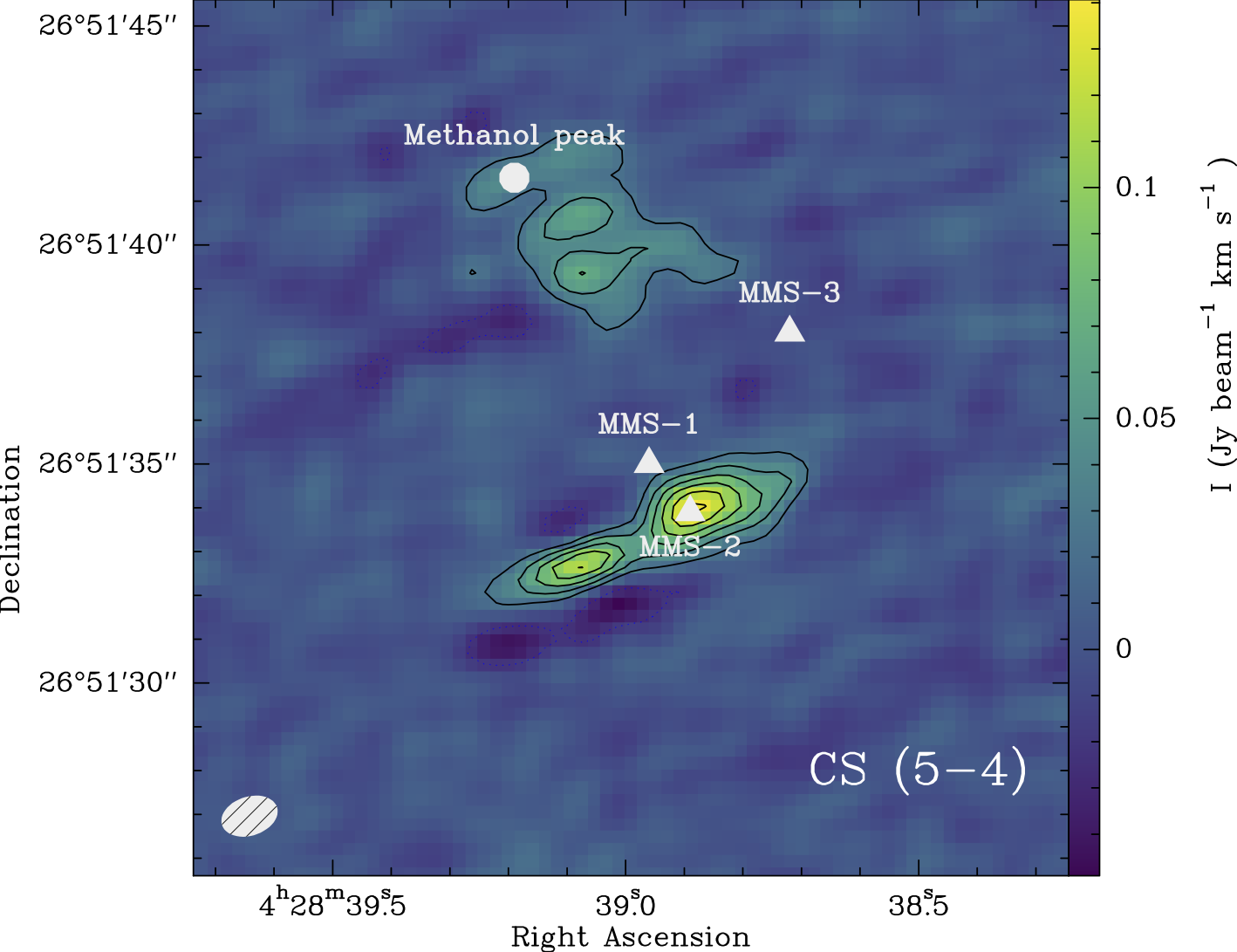}
   \includegraphics[width=0.45\hsize]{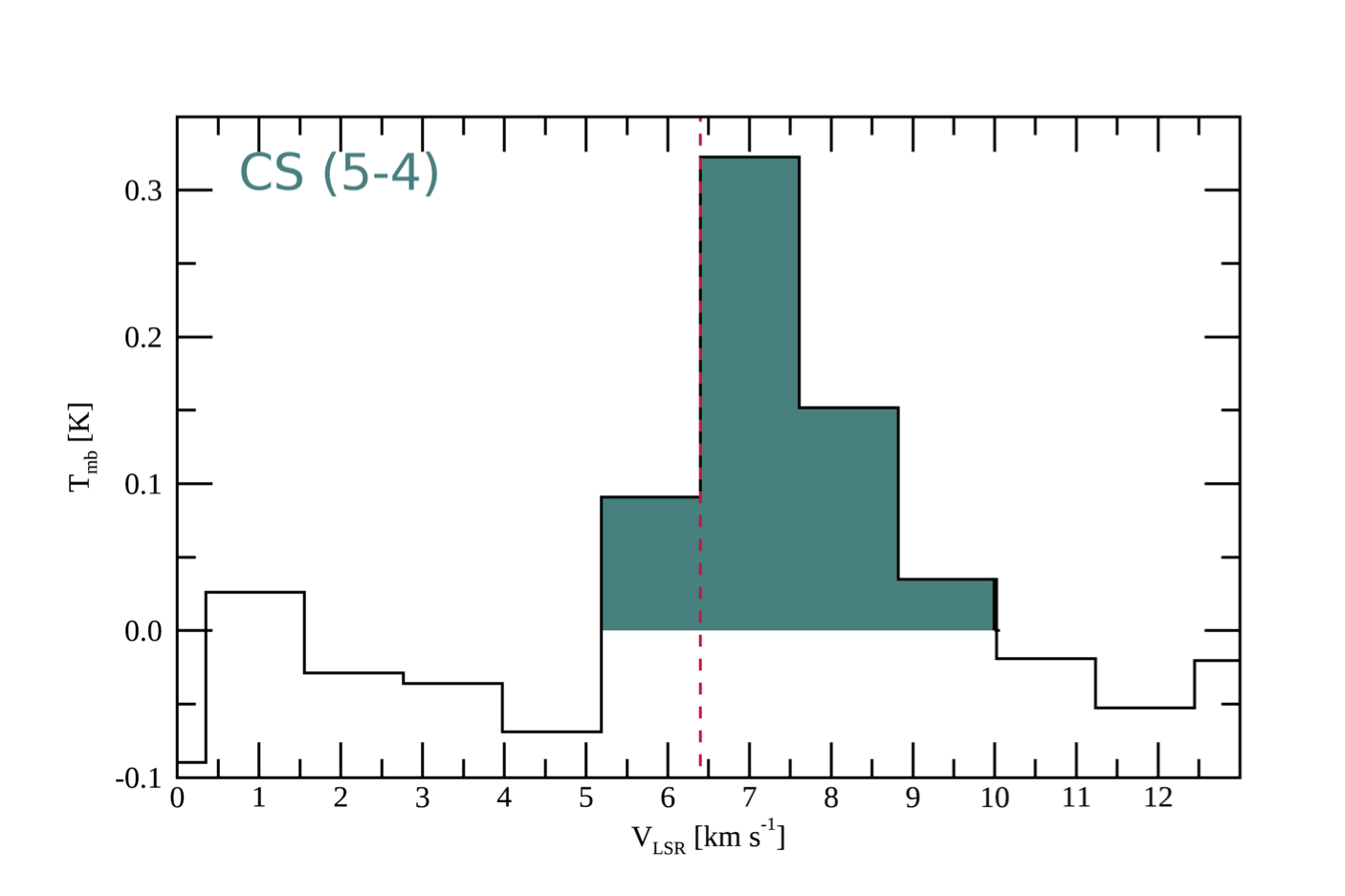}   
  \includegraphics[width=0.45\hsize]{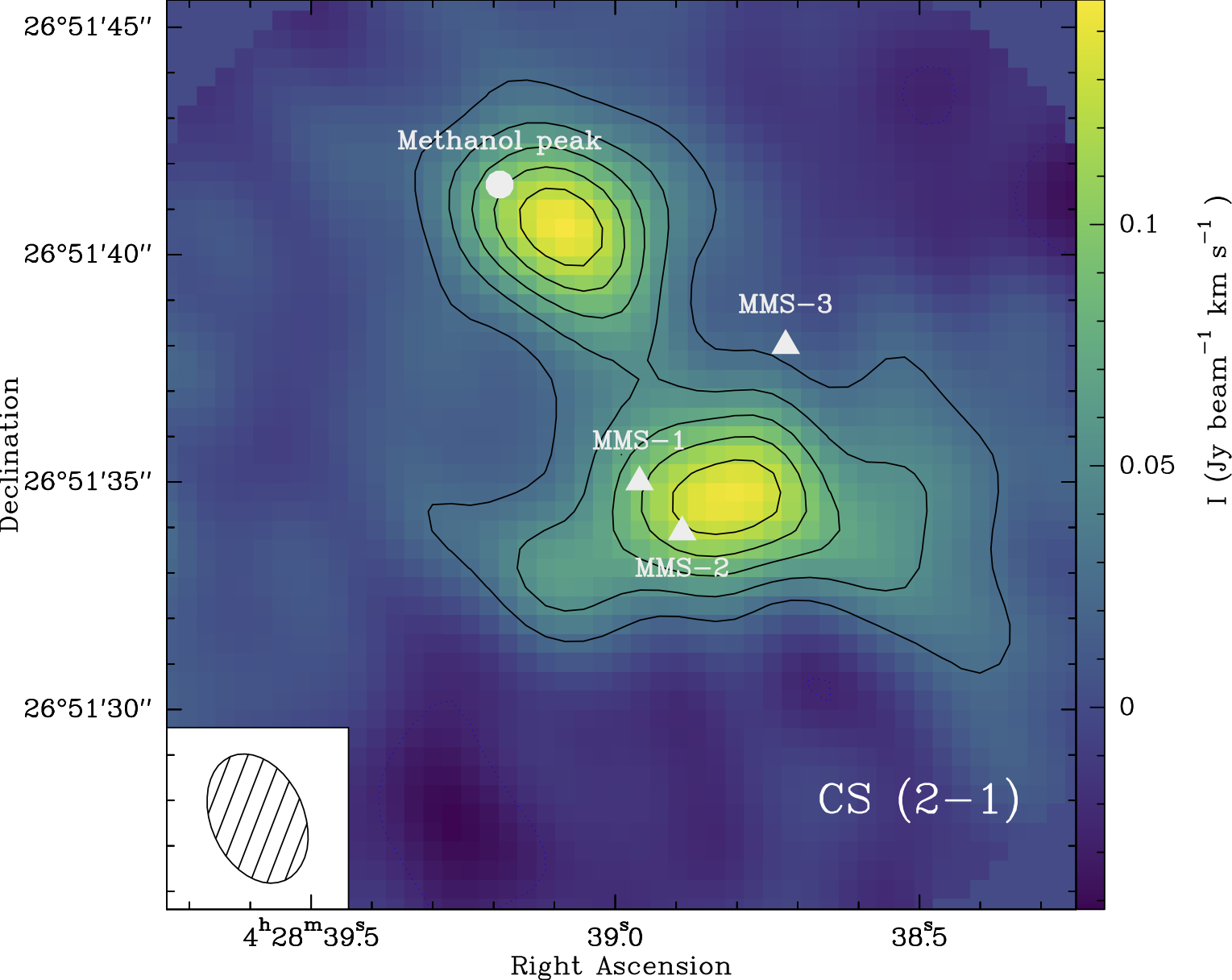}  
    \includegraphics[width=0.45\hsize]{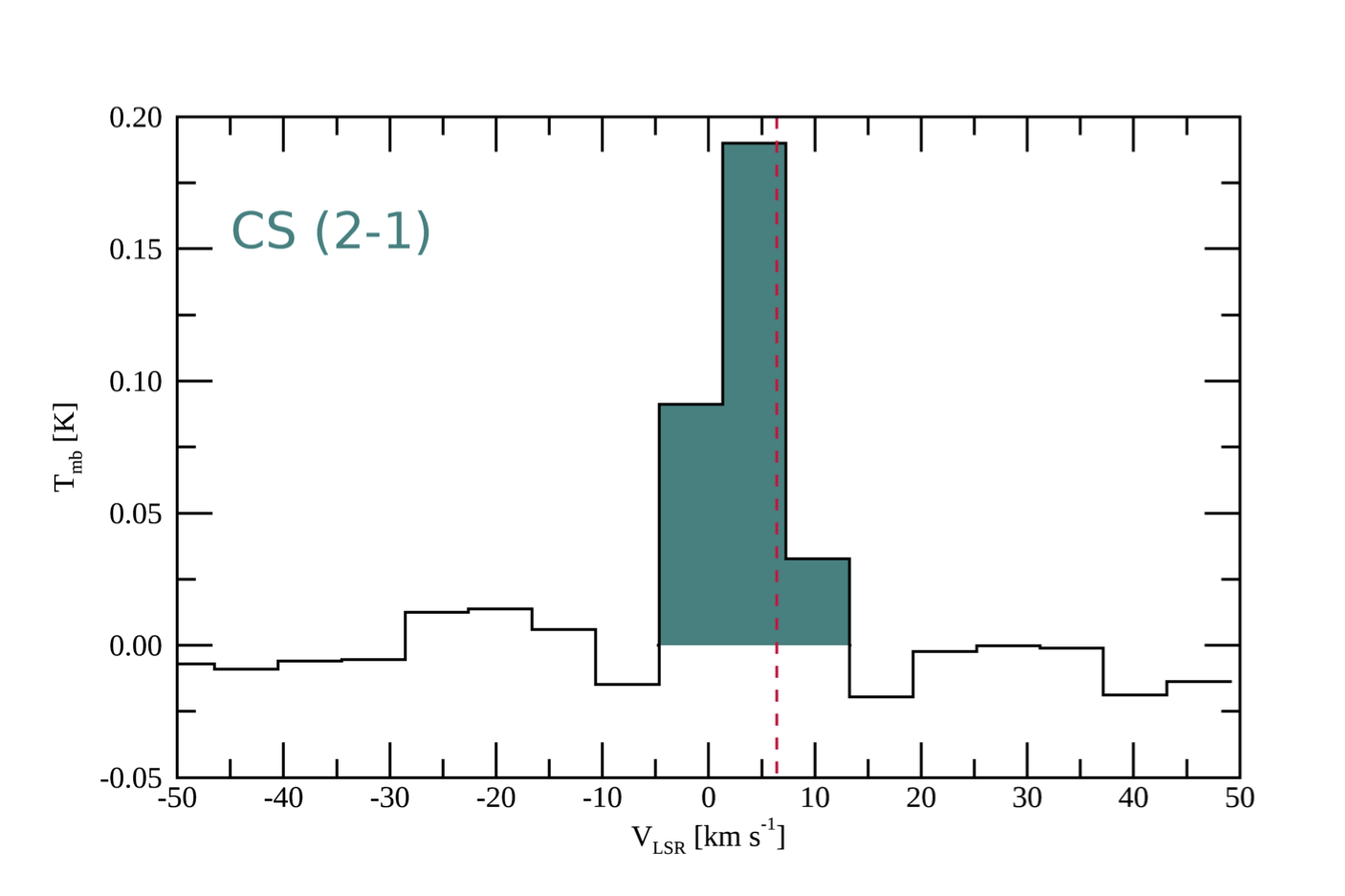}    
  \includegraphics[width=.45\hsize]{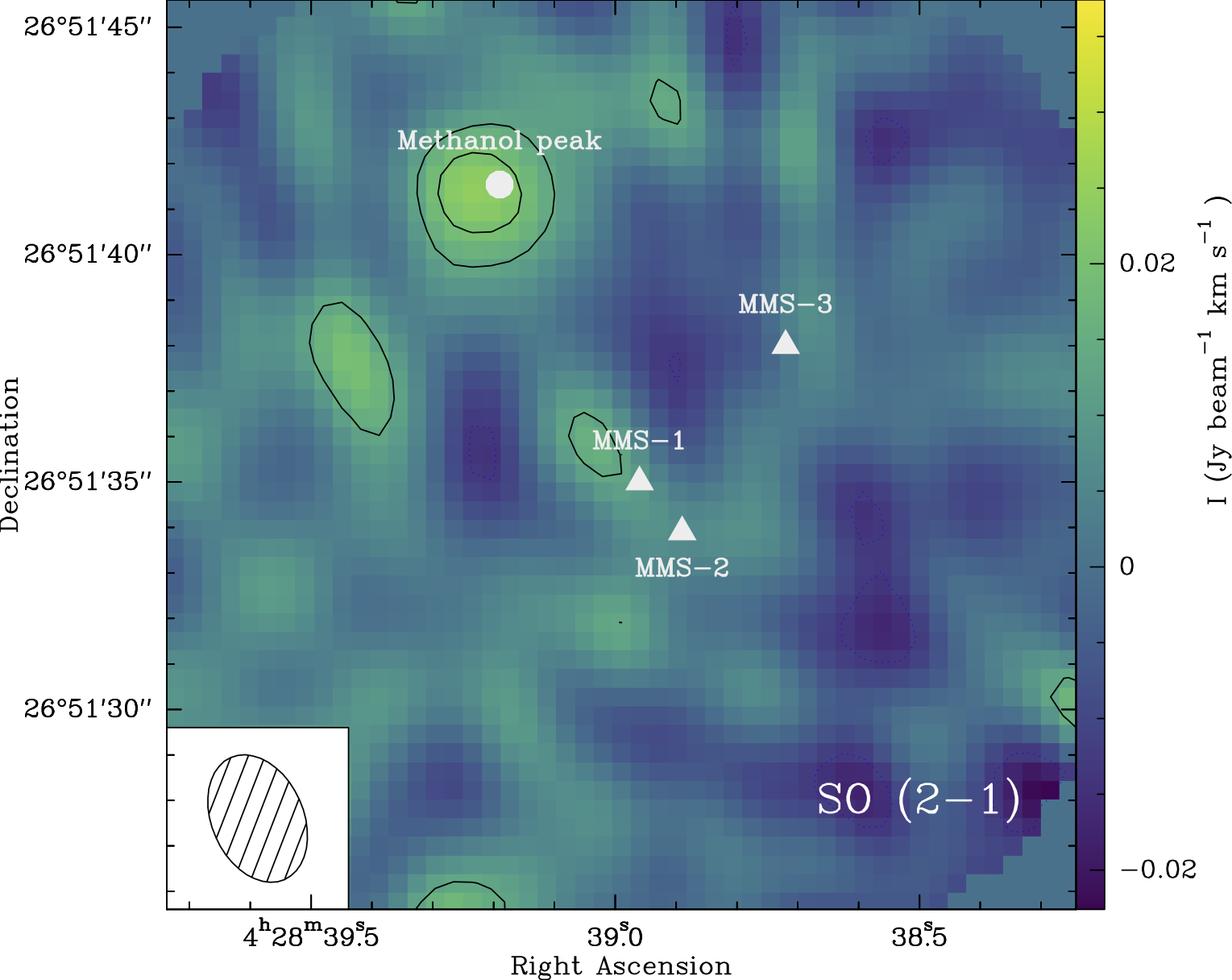} 
  \includegraphics[width=.45\hsize]{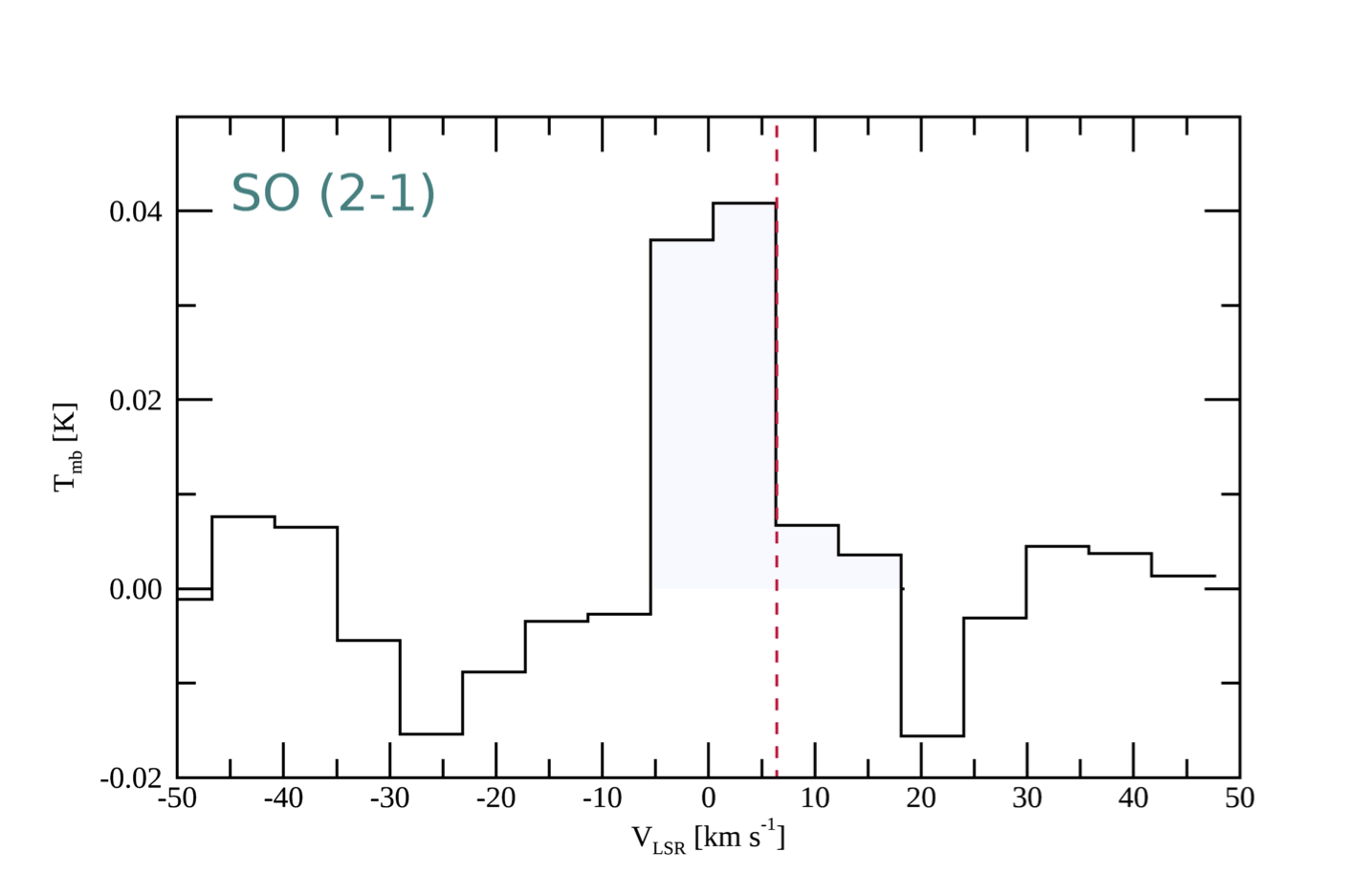}
   \caption{{\it Top left panel:} CS (5-4) integrated intensity emission map as seen with ALMA (project number 2012.1.00239.S, see Section~\ref{Sec:CS}). The contour levels are at  3$\sigma$ (where 1$\sigma$ = 7.5 mJy~beam~$^{-1}$~km~s$^{-1}$). {\it Middle left panel:} CS (2-1) integrated intensity emission map as seen with NOEMA over the line profile (corrected from primary-beam attenuation). The contour levels are at 3$\sigma$ (where 1$\sigma$ = 8.3 mJy~beam~$^{-1}$~km~s$^{-1}$). {\it Bottom left panel:}   SO (2-1) integrated intensity emission map as seen with NOEMA over the line profile. The first contour is at 2$\sigma$ and the level step at 1$\sigma$ (where 1$\sigma$ = 6.6 mJy~beam~$^{-1}$~km~s$^{-1}$). For each map, the synthesized beam is shown in the bottom-left corner. Positions of the sources MMS-1, MMS-2, MMS-3 along with that of the methanol peak (or blob) are indicated. {\it Right panels, from top to bottom:} Spectra of the CS (5-4), CS (2-1) and SO (2-1) spectra taken in direction of the methanol blob. Dashed red lines indicate a V$_{LSR}$ = 6.4~km~s$^{-1}$.}
   \label{fg5}
 \end{figure*}
%+++++++++++++++++++++

\subsection{Distribution of the CS and SO emission}
\label{Sec:CS}

CS (2--1) at 97.980 GHz and SO (2$_3$--1$_2$) at 99.299 GHz have been observed within the WideX bandwidth and the integrated intensity maps are shown in Fig. \ref{fg5}. CS is clearly detected at the position of the methanol blob and at the position of the VeLLO while SO is only tentatively detected toward the methanol fragment. The CS emission is compact ($\sim$ 5$^{\prime\prime}$). In Table \ref{tab1}, we list the spectroscopic and observed parameters of the CS (2--1) line integrated over 5$^{\prime\prime}$. CS (5--4) has also been mapped and detected with ALMA toward L1521F \citep{tokuda2014}. We again retrieved these data from the ALMA data archive (project number 2012.1.00239.S; Early Cycle 0 data performed with a 12m array) and extracted the spectrum within the same circular beam of 5$^{\prime\prime}$ toward the position of the methanol blob (see Table \ref{tab1} for the measured parameters for this line). Both detections indicate that the gas at this position is very dense (see the following section for the radiative transfer).

%+++++++++++++++++++++
%++++ TABLE 1 +++++
\begin{table*}
\caption{Line parameters measured for CS with NOEMA and ALMA toward the position of the methanol peak and integrated over 5$^{\prime\prime}$ using a Gaussian line fitting procedure from the {\sc cassis } software.\label{tab1}}
\center
\begin{tabular}{|c|c|c|c|c|c|c|}
\hline\hline
Species  &  Transition      & Frequency       & v                      & $\Delta$v & Intensity & Integrated Flux\\
               &                        &    (MHz)          & (km~s$^{-1}$)  & (km~s$^{-1}$) & K& (K~km~s$^{-1}$)\\
 \hline
 CS         &   2--1              &   97980.9533   & 5.77(0.24)       & 10.98(0.52) & 0.12(0.01) & 1.32(0.17)\\
 CS         &   5--4              &  244935.5565  & 5.92(0.20)       & 2.86(0.58) & 0.11(0.02) & 0.31(0.12)\\
 \hline
\end{tabular} 
\tablefoot{ {\sc cassis (Centre d'Analyse Scientifique de Spectres Instrumentaux et Synth\'etiques) }: http://cassis.irap.omp.eu. \\ 
The numbers in brackets refer to the 1$\sigma$ level uncertainty derived from Gaussian fit.}
\end{table*} 
%+++++++++++++++++++++

\subsection{Unidentified transition}

\begin{figure}
   \centering
   \includegraphics[width=1\hsize,clip=true,trim=0 0 0 0]{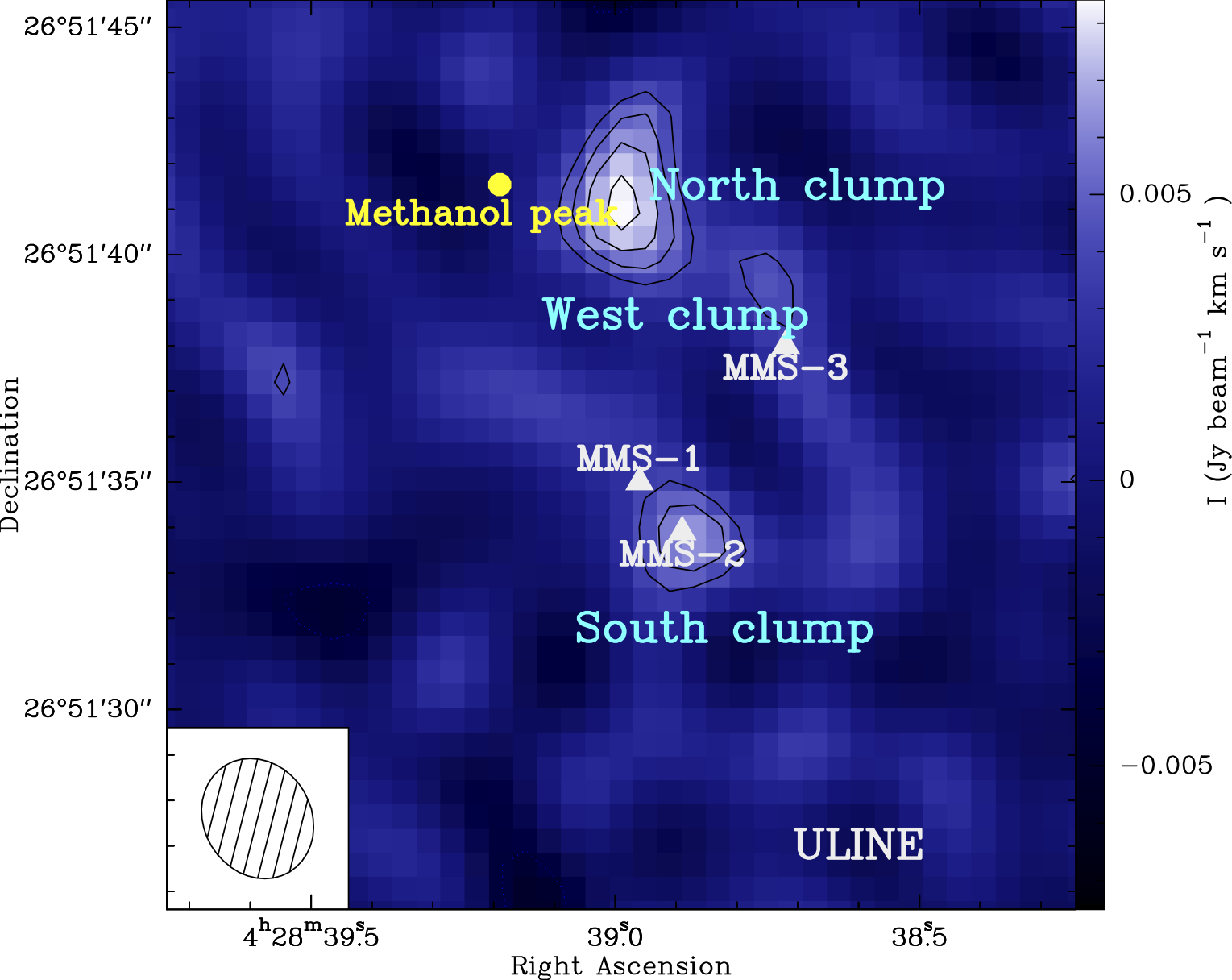}
   \includegraphics[width=1\hsize,clip=true,trim=4cm 0 2.5cm 2.5cm]{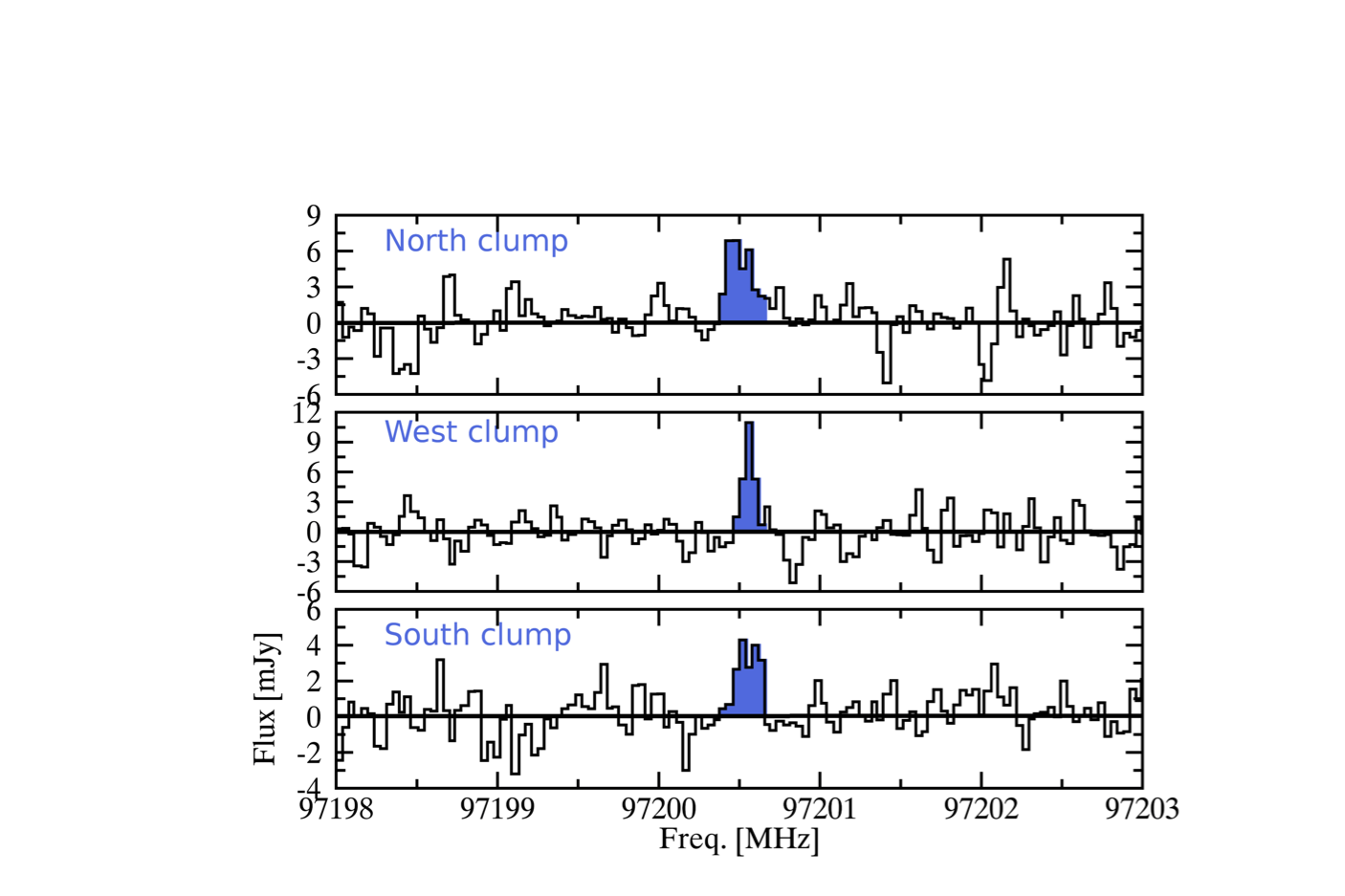}
   \caption{{\it Top panel: }U-line integrated intensity emission map over the line profile. The first contour is at 3$\sigma$ and the level step at 1$\sigma$ (where 1$\sigma$ = 1.3 mJy~beam~$^{-1}$~km~s$^{-1}$). The synthesized beam is shown in the bottom-left corner. Position of the sources MMS-1, MMS-2, MMS-3 along with that of the methanol peak are indicated along with that of the 3 main U-line emission peaks. {\it Bottom panel}: Spectra of the U-line at 97200.541 MHz taken in the direction of the main emission peaks.}
   \label{fg6}
 \end{figure}

Finally we report the detection of an unidentified line (U-line) at the rest frequency of 97200.541 MHz. Three clumps can be resolved in the map seen in Fig. \ref{fg6}: North (corresponding to the methanol peak), West (corresponding to MMS-3) and South (corresponding to MMS-1 and MMS-2). We carefully checked the line identification using the {\sc cassis} software and the CDMS and JPL databases. The line parameters are described in Table \ref{tab2}. The only possible candidate corresponds to vinyl alcohol, a-H$_2$C=CHOH (5$_{3,3}$ -- 4$_{3,2}$, tag=44507 in the CDMS database) with $\nu$= 97200.6 MHz, E$_{up}$ = 36.95 K A$_{ij}$ = 9.3~10$^{-7}$ s$^{-1}$ which has been detected in SgrB2 only by \citet{Turner:2001}. In that instance, the a-H$_2$C=CHOH (2$_{2,1}$ -- 3$_{1,2}$) transition at 96745.9~MHz (with E$_{up}$ = 12.99~K and A$_{ij}$ = 1.1~10$^{-6}$ s$^{-1}$) should also appear in the other NOEMA sub-bands, which is not the case. Therefore, this species can then be dismissed on the basis that one transition only has been detected. We also checked that this U-line is not a remnant from the other sideband. This line does not appear in the IRAM-30m data, probably due to heavy beam dilution. Note that we also explored the L1544 NOEMA data \citep{Punanova:2018} for this transition, which is not detected.

%+++++++++++++++++++++
%++++ FIGURE 6 +++++
%+++++++++++++++++++++

%+++++++++++++++++++++
%++++ TABLE 2 +++++
\begin{table*}
\caption{Location of the U-line and the results from a Gaussian line fitting using the {\sc cassis} software.\label{tab2}}
\center
\begin{tabular}{|c|c|c|c|c|c|}
\hline\hline
 clump                       & ra                     & dec                     &   FWHM              &     Intensity  & rms  \\
                                 &   (J2000)                       &       (J2000)                     &  (km~s$^{-1}$)    &        (mK)        & (mK)\\
\hline
North clump             &  04:28:38.99 & 26:51:41.13  &  0.59 $\pm$ 0.12 &  230  &  62\\
West clump             & 04:28:38.743 & 26:51:39.28 & 0.25 $\pm$ 0.04 &   360 &  64\\
South clump            & 04:28:38.879 & 26:51:33.84 & 0.48 $\pm$ 0.10 & 210 & 56\\
\hline
\end{tabular} 
\end{table*} 
%+++++++++++++++++++++

%===============================================================
%
%-----------------------------------------------------------------------------------------------------------------------------
%------------Molecular column densities and abundances-----------------------------------------------
%-----------------------------------------------------------------------------------------------------------------------------
\section{Molecular column densities and abundances}
\label{sec:abundances}
%-----------------------------------------------------------------------------------------------------------------------------

\subsection{Carbon monosulfide}

CS is an excellent probe of the molecular gas density.  We have used both 2--1 and 5--4 transitions (see Section \ref{sec:results}) to constrain the kinetic temperature and density of the gas in the methanol fragment. The spectral resolution for both transitions is unfortunately very different leading to a much larger linewidth for the 2--1 transition ($\sim$ 11 km~s$^{-1}$, as observed with WideX) compared to the linewidth of $\sim$ 3 km~s$^{-1}$ for the 5--4 transition (measured with ALMA) as shown in Figure~\ref{fg5}. This discrepancy in FHWM is due to the use of data observed with different spectral resolution: the ALMA data were performed with a spectral resolution of 1.4~km~s$^{-1}$ while the NOEMA data (WideX correlator) were observed with a spectral resolution of 6.4~km~s$^{-1}$. We used the collisional rates of CS with H$_2$ calculated by \citet{Lique:2006} for temperatures in the range from 10 to 300 K. 
We first performed a LTE analysis using both {\sc cassis} and {\sc madcuba} \citep{Martin:2019}, assuming that the LTE approximation holds as the derived densities are high. We then, assuming a linewidth of about 3~km~s$^{-1}$, used non-LTE radiative transfer modelling using Radex \citep{van-der-Tak:2007} within {\sc cassis}.
The CS data is consistent with a kinetic temperature range between 10 K and 20 K for a density range [5~10$^5$--3~10$^6$]~cm$^{-3}$ and a column density of [5.5--6.5]~10$^{12}$ cm$^{-2}$. These high densities and cold kinetic temperatures are suggestive of a cold and dense condensation formed within the L1521F star-forming system. 

In a second step we used the line intensity of the central CS (2--1) channel, which has a width of ~3 km~s$^{-1}$, i.e. similar to the CS (5--4) measured linewidth. In that way, we avoid comparing the full integrated intensity of the CS (2--1) lines with the narrower CS (5--4) line. The minimum T$_{k}$ is about 15 K with N(CS)=1.8~10$^{12}$ cm$^{-2}$ and n(H$_2$)>10$^7$ cm$^{-3}$. Using the same calculation but using the line intensity of the CS (2--1) line of the two adjacent channels and the CS (5--4) 3$\sigma$ rms noise level in the spectra, we obtain that N(CS)<8.5~10$^{11}$ cm$^{-2}$ and n(H$_2$)<4~10$^5$ cm$^{-3}$ for the gas at velocities with no CS (5-4) detections. This implies that the density of the methanol blob is a factor 25 higher (at least) than the gas density in the surrounding environment.

\subsection{Methanol}

\begin{figure}
   \includegraphics[width=1.0\hsize]{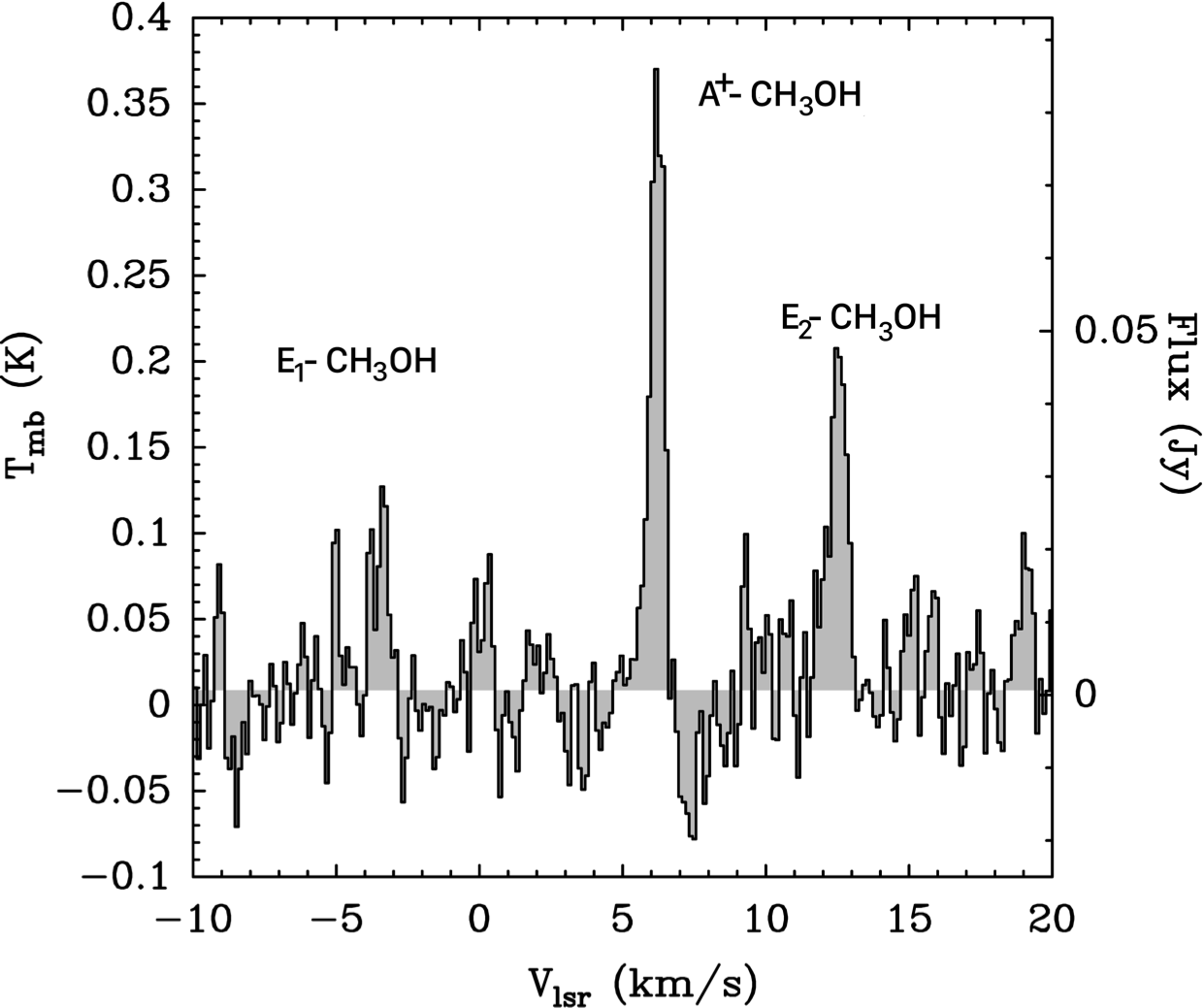}
   \caption{Spectrum of the three methanol transitions in a 7$^{\prime\prime}$ circular beam around the methanol peak/blob using the NOEMA narrow correlator unit. The spectrum is centered at the frequency of the strongest component (A$^+$--CH$_3$OH) at 96.74137~GHz.}
   \label{fg7}
 \end{figure}

 Fig. \ref{fg7} shows the averaged spectrum in a 7$^{\prime\prime}$ beam around the methanol peak/blob using the NOEMA narrow correlator unit (in Kelvin and Jansky). The spectrum is centered at the frequency of the strongest line (A$^+$--CH$_3$OH, with E$_{up}$=6.97 K) at 96.74137~GHz. The E$_2$--CH$_3$OH transition at 96.73936~GHz (E$_{up}$=4.64 K) is also clearly detected at 96.73936 GHz but the 96.74454~GHz E$_1$ transition (E$_{up}$=12.19 K) is marginally detected at the 3 $\sigma$ level with a peak intensity of about 0.1~K. The lines can be fitted with a Gaussian fit with a V$_{LSR}$ of 6.2 km~s$^{-1}$ and a FWHM of $0.60\pm0.05$ km~s$^{-1}$. We present in Table \ref{tab3} the results from the Gaussian line fitting carried out within the {\sc cassis} software. \\

%+++++++++++++++++++++
%++++ TABLE 3 +++++
\begin{table*}
\caption{Line parameters for CH$_3$OH as observed with the narrowband correlator toward the blob integrated over 7$^{\prime\prime}$.  The numbers in brackets refer to the 1$\sigma$ level uncertainty derived from Gaussian fit. \label{tab3}}
\center
\begin{tabular}{|c|c|c|c|c|}
\hline
Frequency  & v & $\Delta$v & Intensity & Integrated Flux \\
 (MHz) & (km~s$^{-1}$)  & (km~s$^{-1}$) & K& (K~km~s$^{-1}$)\\
 \hline
 96739.36  & 6.31(0.04) & 0.63(0.04) & 0.20(0.01) & 0.13(0.02)\\
 96741.37 & 6.21(0.02) & 0.56(0.04) & 0.35(0.02) & 0.20(0.02)\\
 96744.55 & 6.38(0.06) & 0.6(fixed) & $\le$ 0.10 & $\le$ 0.06 (3$\sigma$)\\
 \hline
\end{tabular} 
\end{table*} 
%+++++++++++++++++++++

From the densities derived from the CS transitions (n(H$_2$)>10$^7$ cm$^{-3}$), the methanol lines are most likely in LTE. We performed a simple LTE radiative transfer modelling on the two CH$_3$OH detected transitions as well as on the upper limit, and found an excitation temperature range of (10$\pm$2) K and a column density of (1.3$\pm$0.2)$\times$10$^{13}$ cm$^{-2}$. The resulting excitation temperature is compatible with the kinetic temperature obtained from the non--LTE analysis of the CS transitions. 

%+++++++++++++++++++++
%++++ FIGURE 7 +++++
%+++++++++++++++++++++

%===============================================================
%
%-----------------------------------------------------------------------------------------------------------------------------
%-------------------------------------------- DISCUSSION-------------------------------------------------
%-----------------------------------------------------------------------------------------------------------------------------
\section{Discussion}
\label{sec:discussion}
%-----------------------------------------------------------------------------------------------------------------------------

\subsection{Comparison with previous observations}

The high central density and infall asymmetry seen in the HCO$^+$(3--2) line observed toward L1521F indicate an object in the earliest stages of gravitational collapse \citep{onishi1999}. Detection of a 100 au scale dust-continuum source with 1.3 mm Plateau de Bure Interferometer (PdBI) observations \citep{Maury:2010} supports the claim that the protostar has already formed at the center of L1521F. Single-dish (Caltech Submillimeter Observatory) studies in the CO (7--6 and 6--5) emission detected warm ($\sim$ 30--70 K) and extended ($\sim$ 2400 au) gas, suggesting that this emission may be originating in shocked gas at the interface between the outflow and dense core \citep{shinnaga2009}. 

\citet{tokuda2014} carried out ALMA Cycle 0 observations toward the object at an angular resolution of $\sim$1$^{\prime\prime}$ by using the 12 m array, revealing that the spatial and velocity distributions are very complex. They detected a few starless high-density clumps ( $\sim$10$^7$ cm$^{-3}$), within a region of several hundred au around the Spitzer source, a very compact bipolar outflow centered at the protostar source with a dynamical time of a few hundred years with an indication of interaction of surrounding gas and a well-defined long arc-like structure whose size is  $\sim$ 2000 au. More recent ALMA Cycle 1 observations have been performed by \citet{Tokuda:2016} with a sub-arcsecond resolution, leading to the detection of three intensity peaks at 0.87 mm (MMS-1: ${\rm \alpha_{2000} = 04^h28^m38.96^s, \delta_{2000} = 26\degr51\arcmin35\arcsec}$, MMS-2: ${\rm \alpha_{2000} = 04^h28^m38.89^s, \delta_{2000} = 26\degr51\arcmin33.9\arcsec}$ and MMS-3: ${\rm \alpha_{2000} = 04^h28^m38.72^s, \delta_{2000} = 26\degr51\arcmin38\arcsec}$). MMS-1 corresponds to the Spitzer source, L1521F-IRS. Their CO (3--2) and HCO$^+$ (3--2) observations reveal a complex structure that links all three 0.87 mm peaks as well as the L1521F-NE source detected by \citet{takahashi2013} with the SMA interferometer. The CO blueshifted and redshifted components observed by the SMA, are distributed symmetrically and seem to result from multiple outflows from a binary system, one associated with L1521F-IRS and another one associated with this new source, L1521F-NE. However, no driving source has been detected in either millimeter continuum emission with PdBI/SMA or infrared emission with Spitzer. Note that \citet{Tokuda:2016} do not believe in the outflows identified by \citet{takahashi2013} since the direction of the outflow is inconsistent with the Spitzer reflection nebula and suggests instead that it is a relatively high-density gas structure surrounding L1521F.  The molecular line observation showed several cores with arc-like structures, possibly due to the dynamical gas interaction. Similar arc-like structures have been reproduced by hydrodynamical simulations with and without magnetic field \citep{Matsumoto:2015,Matsumoto:2017}. The complex structure indicates that in this source turbulence, probably injected by the protostellar feedback, may play an essential role in undergoing fragmentation in the central part of the cloud core. The mechanism is different from the classic scenarios of fragmentation in massive disks \citep{Larson:1987,Boss:2002,Machida:2008}. All the above single dish and interferometric observations demonstrated that significant temperature variations are identified within the core, justifying the need for a high spatial resolution of the central regions.\\

\subsection{The CH$_3$OH ring-like structure in L1521F}

The CH$_3$OH emission peak detected in the L1521F cluster apears at a distance about 1000 au from the center of the core. In an analogy to the L1544 prestellar core, the CH$_3$OH peak found with NOEMA toward L1521F resembles the CH$_3$OH peak reported at $\sim$4000 au toward the north-east of L1544 \citep{bizzocchi2014}. In addition, as shown in Figure$\,$\ref{fg4}, L1521F also shows a ring-like structure in CH$_3$OH around the MMS-1 source, which is also similar to that observed toward the L1544 prestellar core \citep{bizzocchi2014}. The CH$_3$OH peak and ring in L1544 coincides with the region in the core where CO starts to freeze out and deuterium fractionation starts to be enhanced \citep{Caselli:1999,Caselli:2002}.

This ring-like morphology of the CH$_3$OH emission in L1521F and L1544 has also been reported in other prestellar cores \citep[see also][]{tafalla2004}, and it is likely the result of several factors: i) the depletion of C-bearing species as the density increases with decreasing radii within the core; ii) non-thermal desorption processes such as chemical reactive desorption;  iii) the photo-destruction of CH$_3$OH at visual extinctions $A_v$$\leq$5 mag in the outskirts of the core \citep{vasyunin2017}; iv) sputtering from a gentle shock (see Section 5.3). As a consequence, it is expected that molecular complexity is luxuriant in the external layers of prestellar cores, as confirmed observationally in L1544 \citep{Vastel:2014,vastel2016,vastel2018b,vastel2019,jimenez-serra2016,quenard2017a}. In fact, at the location of the methanol peak, \citet{jimenez-serra2016} found oxygen-bearing complex organic molecules such as CH$_3$CHO, HCOOCH$_3$ and CH$_3$OCH$_3$, as well as methoxy (CH$_3$O), all related to the release of methanol in the gas phase \citep{Balucani:2015,Soma:2015,Bertin2016,vasyunin2017}. 

\subsection{On the nature of the CH$_3$OH blob in L1521F}

The NOEMA-only maps of CH$_3$OH obtained toward L1521F, resolve out extended methanol emission and reveal that the methanol peak/blob is very compact (950 au). The physical properties derived for this blob are T$_k$ $\sim$ (10$\pm$2) K and n(H$_2$)>10$^7$ cm$^{-3}$. If we compare these values to those of the L1544 prototypical prestellar core (i.e. n(H$_2$) $\sim$ 10$^6$ cm$^{-3}$ and T$_k$ $\sim$ 7 K for a 500 au radius; see \citealp{Crapsi:2007} and Fig. 2 in \citealp{vastel2018b}), we find that they are quite similar. However, note that while the methanol peak in L1544 is located at $\sim$4000 au from the core center, the methanol blob in L1521F is found at roughly $\sim$1000 au. At this distance, methanol is clearly depleted in L1544 \citep{bizzocchi2014,Vastel:2014,jimenez-serra2014,Punanova:2018} due to the high density and low temperatures found at this distance in the core. 

 \citet{crapsi2004} used the 1.2 mm continuum data of L1521F from the IRAM-30m to estimate the density distribution under the assumption of spherical symmetry. They followed the same technique adopted by \citet{Tafalla:2002} and best-fitted their data with a model of the form:\\
\begin{equation}
n(r) = \frac{10^6}{1+\left(\frac{r}{20^{\prime\prime}}\right)^2}~.
\end{equation}
With this density profile, the expected density at the location of the methanol blob is about 9$\times$10$^5$ cm$^{-3}$, lower than the density estimated in Section 4.2 (larger than 10$^{7}$ cm$^{-3}$) from the excitation analysis of the CS (2-1) and (5-4) lines. 
The resulting derived density appears to be higher than that expected from the n(H$_2$) gas density distribution; suggesting that the CH$_3$OH blob might has undergone a compression event of some sort. \\

In that context, recent NOEMA observations of L1544 within the SOLIS Large program, focused on the small-scale morphology of the methanol peak emission \citep{Punanova:2018}. The kinetic temperature and H$_2$ gas column density measured for the methanol peak in L1544 from the NOEMA data are 10~K and (2.3$\pm$0.3) $\times$10$^{22}$~cm$^{-3}$. \citet{Punanova:2018} concluded that this local methanol enhancement could be an indication of gentle accretion of material onto the core or the interaction of two filaments producing a slow shock. The methanol peak emission in L1544 is much more extended (more than 20$^{\prime\prime}$) that the peak emission detected in the L1521F region ($\sim$ 5$^{\prime\prime}$), and appears closer to the center of the core.
It is interesting to note that, no thermal continuum emission is detected toward the methanol blob. Therefore, the fragmentation scenario (see Section~5.1), could be a possible explanation for our observations. \\

Finally, as briefly mentioned in the previous section, the very presence of methanol in the gas phase in such a cold ($\sim$10 K) environment is itself a strong message on its origin. Specifically, since methanol is believed to be a grain-surface product \citep[e.g.][]{Watanabe:2002,Rimola:2014} and the temperature is too low for thermal desorption to play a role, some other mechanism is at work. A first one could be the photodesorption from UV photons penetrating up to the methanol blob but laboratory experiments suggest that the iced methanol would be injected into the gas phase only as fragments (such as CH$_3$O) and not as whole molecules \citep{Bertin2016}. A second, often evoked mechanism is the so-called chemical desorption, namely the idea that the energy of the grain-surface reaction is partially transmitted to the product, in this case methanol, to desorb it. While this mechanism could be valid for some species \citep[see e.g.][]{Oba:2018}, it does not seem efficient for methanol, based on laboratory experiments \citep{Minissale:2016,Chuang:2018}. However, if the composition of the icy mantles includes a larger concentration of CO, methanol could be efficiently chemically desorbed \citep[see][]{vasyunin2017}. 
A last possibility is represented by the sputtering caused by a gentle shock \citep[e.g.][]{Flower:1995}. This last hypothesis seems to be the most probable for the following reasons: (i) the location of the methanol blob, not coinciding with any continuum emission peak; (ii) the high density ($\geq 10^7$ cm$^{-3}$), higher than the surrounding one by more than a factor 25; (iii) the relatively small extent ($\sim5^{\prime\prime}$), indicating a very localized phenomenon. If the methanol blob is due to such a gentle shock, then also the presence of methanol in the gas-phase would be easier to explain.\\
If the methanol blob is due to a gentle shock then the latter is extremely recent, as methanol would freeze out back onto the grain mantles very quickly, in a few hundred years. This could also explain why SO is not detected in our observations. If SO, as commonly assumed, is formed in the gas-phase by oxidation of sulfur released from the grain mantles in the form of S or other hydrogenated/organo/metallic S-bearing species \citep[e.g.][]{Laas:2019}, SO would need a few thousands years to form \citep[depending on the gas temperature history: e.g.][]{Wakelam:2004,Wakelam:2018}.\\

In summary, the hypothesis that the methanol blob is recent, $\leq$ few hundred years, a shock would likely explain the  presence of methanol and the absence of SO in the gas-phase. The origin of this shock could be a channel of infalling gas toward the center of L1521F. Alternatively, we cannot exclude the hypothesis of the formation of a cold and dense methanol fragment due to gas dynamics.

%===============================================================
%
%-----------------------------------------------------------------------------------------------------------------------------
%-------------------------------------------- CONCLUSIONS------------------------------------------------
%-----------------------------------------------------------------------------------------------------------------------------
\section{Conclusions}

The original goal of the SOLIS IRAM-NOEMA Large Program to detect several crucial organic molecules in a sample of Solar-like star forming regions in different evolutionary stages and environments is not achieved for the L1521F very low luminosity object. Instead, we revealed for the first time the presence of a methanol blob emission in the North-East part of the region, which is located at about $\sim$1000 au away from the L1521F source. Our study suggests that we are observing, at the intersection of a filamentary system (studied previously with ALMA in HCO$^+$ and CO) either the formation {\it i)} of a shock-induced cold dense blob or {\it ii)} that of a cold dense fragment. Further observations are needed to distinguish between the two scenarios. \\

Finally, these observations took place before the implementation of the wideband high-performance correlator PolyFiX that achieved a much higher sensitivity and a much larger bandwidth. A follow-up study at the IRAM 30m will be presented in a forthcoming paper, with deuterated species as well as COMs. \\

%=============================================
%
%-----------------------------------------------------------------
%----------ACKNOWLEDGEMENTS-----
%-----------------------------------------------------------------
\begin{acknowledgements}
We thank our referee, Dr. Kazuki Tokuda, {\it i)} for his fruitful comments that have improved the quality of our paper and  {\it ii)} for sharing his continuum emission map. This work is supported by the French National Research Agency in the framework of the Investissements d'Avenir program (ANR-15-IDEX-02), through the funding of the "Origin of Life" project of the Univ. Grenoble-Alpes. 
CF, CV and CC acknowledge the funding from the European Research Council (ERC) under the European Unions Horizon 2020 research and innovation programme, for the Project {\it The Dawn of Organic Chemistry} (DOC), grant agreement No 741002. 
I.J.-S. has received partial support from the Spanish FEDER (project number ESP2017-86582-C4-1-R), and State Research Agency (AEI) through project number MDM-2017-0737 Unidad de Excelencia {\it Mar\'ia de Maeztu}--Centro de Astrobiolog\'ia (INTA-CSIC).
AP acknowledges the financial support of the Russian Science Foundation project 18--12--00351.
A.C.-T acknowledges support from MINECO project AYA2016-79006-P.
\end{acknowledgements}

%===============================================================
%
%-----------------------------------------------------------------------------------------------------------------------------
%-----------------------------------------BIBLIO----------------------------------
%-----------------------------------------------------------------------------------------------------------------------------
%
\bibliographystyle{aa}
%\bibliography{/Users/favrecec/Documents/Articles/biblio}

\begin{thebibliography}{76}
\expandafter\ifx\csname natexlab\endcsname\relax\def\natexlab#1{#1}\fi

\bibitem[{{Andr{\'e}} {et~al.}(2019){Andr{\'e}}, {Arzoumanian}, {K{\"o}nyves},
  {Shimajiri}, \& {Palmeirim}}]{Andre:2019}
{Andr{\'e}}, P., {Arzoumanian}, D., {K{\"o}nyves}, V., {Shimajiri}, Y., \&
  {Palmeirim}, P. 2019, \aap, 629, L4

\bibitem[{{Andr{\'e}} {et~al.}(2010){Andr{\'e}}, {Men'shchikov}, {Bontemps},
  {K{\"o}nyves}, {Motte}, {Schneider}, {Didelon}, {Minier}, {Saraceno},
  {Ward-Thompson}, {di Francesco}, {White}, {Molinari}, {Testi}, {Abergel},
  {Griffin}, {Henning}, {Royer}, {Mer{\'\i}n}, {Vavrek}, {Attard},
  {Arzoumanian}, {Wilson}, {Ade}, {Aussel}, {Baluteau}, {Benedettini},
  {Bernard}, {Blommaert}, {Cambr{\'e}sy}, {Cox}, {di Giorgio}, {Hargrave},
  {Hennemann}, {Huang}, {Kirk}, {Krause}, {Launhardt}, {Leeks}, {Le Pennec},
  {Li}, {Martin}, {Maury}, {Olofsson}, {Omont}, {Peretto}, {Pezzuto}, {Prusti},
  {Roussel}, {Russeil}, {Sauvage}, {Sibthorpe}, {Sicilia-Aguilar}, {Spinoglio},
  {Waelkens}, {Woodcraft}, \& {Zavagno}}]{Andre:2010}
{Andr{\'e}}, P., {Men'shchikov}, A., {Bontemps}, S., {et~al.} 2010, \aap, 518,
  L102

\bibitem[{{Balucani} {et~al.}(2015){Balucani}, {Ceccarelli}, \&
  {Taquet}}]{Balucani:2015}
{Balucani}, N., {Ceccarelli}, C., \& {Taquet}, V. 2015, \mnras, 449, L16

\bibitem[{{Bertin} {et~al.}(2016){Bertin}, {Romanzin}, {Doronin}, {Philippe},
  {Jeseck}, {Ligterink}, {Linnartz}, {Michaut}, \& {Fillion}}]{Bertin2016}
{Bertin}, M., {Romanzin}, C., {Doronin}, M., {et~al.} 2016, ApJL, 817, L12

\bibitem[{{Bizzocchi} {et~al.}(2014){Bizzocchi}, {Caselli}, {Spezzano}, \&
  {Leonardo}}]{bizzocchi2014}
{Bizzocchi}, L., {Caselli}, P., {Spezzano}, S., \& {Leonardo}, E. 2014, \aap,
  569, A27

\bibitem[{{Boss}(2002)}]{Boss:2002}
{Boss}, A.~P. 2002, \apj, 568, 743

\bibitem[{{Bourke} {et~al.}(2006){Bourke}, {Myers}, {Evans}, {Dunham},
  {Kauffmann}, {Shirley}, {Crapsi}, {Young}, {Huard}, {Brooke}, {Chapman},
  {Cieza}, {Lee}, {Teuben}, \& {Wahhaj}}]{bourke2006}
{Bourke}, T.~L., {Myers}, P.~C., {Evans}, Neal~J., I., {et~al.} 2006, \apjl,
  649, L37

\bibitem[{{Caselli} {et~al.}(1999){Caselli}, {Walmsley}, {Tafalla}, {Dore}, \&
  {Myers}}]{Caselli:1999}
{Caselli}, P., {Walmsley}, C.~M., {Tafalla}, M., {Dore}, L., \& {Myers}, P.~C.
  1999, \apjl, 523, L165

\bibitem[{{Caselli} {et~al.}(2002){Caselli}, {Walmsley}, {Zucconi}, {Tafalla},
  {Dore}, \& {Myers}}]{Caselli:2002}
{Caselli}, P., {Walmsley}, C.~M., {Zucconi}, A., {et~al.} 2002, \apj, 565, 344

\bibitem[{{Ceccarelli} {et~al.}(2017){Ceccarelli}, {Caselli}, {Fontani},
  {Neri}, {L{\'o}pez-Sepulcre}, {Codella}, {Feng}, {Jim{\'e}nez-Serra},
  {Lefloch}, {Pineda}, {Vastel}, {Alves}, {Bachiller}, {Balucani}, {Bianchi},
  {Bizzocchi}, {Bottinelli}, {Caux}, {Chac{\'o}n-Tanarro}, {Choudhury},
  {Coutens}, {Dulieu}, {Favre}, {Hily-Blant}, {Holdship}, {Kahane}, {Jaber
  Al-Edhari}, {Laas}, {Ospina}, {Oya}, {Podio}, {Pon}, {Punanova}, {Quenard},
  {Rimola}, {Sakai}, {Sims}, {Spezzano}, {Taquet}, {Testi}, {Theul{\'e}},
  {Ugliengo}, {Vasyunin}, {Viti}, {Wiesenfeld}, \&
  {Yamamoto}}]{Ceccarelli:2017}
{Ceccarelli}, C., {Caselli}, P., {Fontani}, F., {et~al.} 2017, \apj, 850, 176

\bibitem[{{Chen} {et~al.}(2012){Chen}, {Arce}, {Dunham}, {Zhang}, {Bourke},
  {Launhardt}, {Schmalzl}, \& {Henning}}]{Chen:2012}
{Chen}, X., {Arce}, H.~G., {Dunham}, M.~M., {et~al.} 2012, \apj, 751, 89

\bibitem[{{Chen} {et~al.}(2010){Chen}, {Arce}, {Zhang}, {Bourke}, {Launhardt},
  {Schmalzl}, \& {Henning}}]{Chen:2010}
{Chen}, X., {Arce}, H.~G., {Zhang}, Q., {et~al.} 2010, \apj, 715, 1344

\bibitem[{{Chuang} {et~al.}(2018){Chuang}, {Fedoseev}, {Qasim}, {Ioppolo}, {van
  Dishoeck}, \& {Linnartz}}]{Chuang:2018}
{Chuang}, K.~J., {Fedoseev}, G., {Qasim}, D., {et~al.} 2018, \aap, 617, A87

\bibitem[{{Codella} {et~al.}(1997){Codella}, {Welser}, {Henkel}, {Benson}, \&
  {Myers}}]{codella1997}
{Codella}, C., {Welser}, R., {Henkel}, C., {Benson}, P.~J., \& {Myers}, P.~C.
  1997, \aap, 324, 203

\bibitem[{{Commer{\c{c}}on} {et~al.}(2012){Commer{\c{c}}on}, {Launhardt},
  {Dullemond}, \& {Henning}}]{Commercon:2012}
{Commer{\c{c}}on}, B., {Launhardt}, R., {Dullemond}, C., \& {Henning}, T. 2012,
  \aap, 545, A98

\bibitem[{{Crapsi} {et~al.}(2005){Crapsi}, {Caselli}, {Walmsley}, {Myers},
  {Tafalla}, {Lee}, \& {Bourke}}]{crapsi2005}
{Crapsi}, A., {Caselli}, P., {Walmsley}, C.~M., {et~al.} 2005, \apj, 619, 379

\bibitem[{{Crapsi} {et~al.}(2004){Crapsi}, {Caselli}, {Walmsley}, {Tafalla},
  {Lee}, {Bourke}, \& {Myers}}]{crapsi2004}
{Crapsi}, A., {Caselli}, P., {Walmsley}, C.~M., {et~al.} 2004, \aap, 420, 957

\bibitem[{{Crapsi} {et~al.}(2007){Crapsi}, {Caselli}, {Walmsley}, \&
  {Tafalla}}]{Crapsi:2007}
{Crapsi}, A., {Caselli}, P., {Walmsley}, M.~C., \& {Tafalla}, M. 2007, \aap,
  470, 221

\bibitem[{{Dunham} {et~al.}(2011){Dunham}, {Chen}, {Arce}, {Bourke}, {Schnee},
  \& {Enoch}}]{Dunham:2011}
{Dunham}, M.~M., {Chen}, X., {Arce}, H.~G., {et~al.} 2011, \apj, 742, 1

\bibitem[{{Dunham} {et~al.}(2008){Dunham}, {Crapsi}, {Evans}, {Bourke},
  {Huard}, {Myers}, \& {Kauffmann}}]{Dunham:2008}
{Dunham}, M.~M., {Crapsi}, A., {Evans}, Neal~J., I., {et~al.} 2008, \apjs, 179,
  249

\bibitem[{{Dunham} {et~al.}(2006){Dunham}, {Evans}, {Bourke}, {Dullemond},
  {Young}, {Brooke}, {Chapman}, {Myers}, {Porras}, {Spiesman}, {Teuben}, \&
  {Wahhaj}}]{Dunham:2006}
{Dunham}, M.~M., {Evans}, Neal~J., I., {Bourke}, T.~L., {et~al.} 2006, \apj,
  651, 945

\bibitem[{{Enoch} {et~al.}(2010){Enoch}, {Lee}, {Harvey}, {Dunham}, \&
  {Schnee}}]{Enoch:2010}
{Enoch}, M.~L., {Lee}, J.-E., {Harvey}, P., {Dunham}, M.~M., \& {Schnee}, S.
  2010, \apjl, 722, L33

\bibitem[{{Flower} \& {Pineau des Forets}(1995)}]{Flower:1995}
{Flower}, D.~R. \& {Pineau des Forets}, G. 1995, \mnras, 275, 1049

\bibitem[{{Hosokawa} {et~al.}(2011){Hosokawa}, {Omukai}, {Yoshida}, \&
  {Yorke}}]{Hosokawa:2011}
{Hosokawa}, T., {Omukai}, K., {Yoshida}, N., \& {Yorke}, H.~W. 2011, Science,
  334, 1250

\bibitem[{{Jim{\'e}nez-Serra} {et~al.}(2014){Jim{\'e}nez-Serra}, {testi},
  {Caselli}, \& {Viti}}]{jimenez-serra2014}
{Jim{\'e}nez-Serra}, I., {testi}, L., {Caselli}, P., \& {Viti}, S. 2014, \apjl,
  787, L33

\bibitem[{{Jim{\'e}nez-Serra} {et~al.}(2016){Jim{\'e}nez-Serra}, {Vasyunin},
  {Caselli}, {Marcelino}, {Billot}, {Viti}, {Testi}, {Vastel}, {Lefloch}, \&
  {Bachiller}}]{jimenez-serra2016}
{Jim{\'e}nez-Serra}, I., {Vasyunin}, A.~I., {Caselli}, P., {et~al.} 2016,
  \apjl, 830, L6

\bibitem[{{Laas} \& {Caselli}(2019)}]{Laas:2019}
{Laas}, J.~C. \& {Caselli}, P. 2019, \aap, 624, A108

\bibitem[{{Larson}(1969)}]{Larson:1969}
{Larson}, R.~B. 1969, \mnras, 145, 271

\bibitem[{{Larson}(1987)}]{Larson:1987}
{Larson}, R.~B. 1987, American Scientist, 75, 376

\bibitem[{{Lee} {et~al.}(2009){Lee}, {Bourke}, {Myers}, {Dunham}, {Evans},
  {Lee}, {Huard}, {Wu}, {Gutermuth}, {Kim}, \& {Kang}}]{Lee:2009a}
{Lee}, C.~W., {Bourke}, T.~L., {Myers}, P.~C., {et~al.} 2009, \apj, 693, 1290

\bibitem[{{Lique} {et~al.}(2006){Lique}, {Spielfiedel}, \&
  {Cernicharo}}]{Lique:2006}
{Lique}, F., {Spielfiedel}, A., \& {Cernicharo}, J. 2006, \aap, 451, 1125

\bibitem[{{Machida} {et~al.}(2008){Machida}, {Inutsuka}, \&
  {Matsumoto}}]{Machida:2008}
{Machida}, M.~N., {Inutsuka}, S.-i., \& {Matsumoto}, T. 2008, \apj, 676, 1088

\bibitem[{{Maheswar} {et~al.}(2011){Maheswar}, {Lee}, \& {Dib}}]{Maheswar:2011}
{Maheswar}, G., {Lee}, C.~W., \& {Dib}, S. 2011, \aap, 536, A99

\bibitem[{{Mart{\'\i}n} {et~al.}(2019){Mart{\'\i}n}, {Mart{\'\i}n-Pintado},
  {Blanco-S{\'a}nchez}, {Rivilla}, {Rodr{\'\i}guez-Franco}, \&
  {Rico-Villas}}]{Martin:2019}
{Mart{\'\i}n}, S., {Mart{\'\i}n-Pintado}, J., {Blanco-S{\'a}nchez}, C.,
  {et~al.} 2019, \aap, 631, A159

\bibitem[{{Masunaga} \& {Inutsuka}(2000)}]{Masunaga:2000}
{Masunaga}, H. \& {Inutsuka}, S.-i. 2000, \apj, 531, 350

\bibitem[{{Masunaga} {et~al.}(1998){Masunaga}, {Miyama}, \&
  {Inutsuka}}]{Masunaga:1998}
{Masunaga}, H., {Miyama}, S.~M., \& {Inutsuka}, S.-i. 1998, \apj, 495, 346

\bibitem[{{Matsumoto} {et~al.}(2015){Matsumoto}, {Onishi}, {Tokuda}, \&
  {Inutsuka}}]{Matsumoto:2015}
{Matsumoto}, T., {Onishi}, T., {Tokuda}, K., \& {Inutsuka}, S.~I. 2015, \mnras,
  449, L123

\bibitem[{{Matsumoto} {et~al.}(2017){Matsumoto}, {Tokuda}, {Onishi},
  {Inutsuka}, {Saigo}, \& {Takakuwa}}]{Matsumoto:2017}
{Matsumoto}, T., {Tokuda}, K., {Onishi}, T., {et~al.} 2017, in Journal of
  Physics Conference Series, Vol. 837, 012009

\bibitem[{{Maury} {et~al.}(2010){Maury}, {Andr{\'e}}, {Hennebelle}, {Motte},
  {Stamatellos}, {Bate}, {Belloche}, {Duch{\^e}ne}, \&
  {Whitworth}}]{Maury:2010}
{Maury}, A.~J., {Andr{\'e}}, P., {Hennebelle}, P., {et~al.} 2010, \aap, 512,
  A40

\bibitem[{{Maury} {et~al.}(2019){Maury}, {Andr{\'e}}, {Testi}, {Maret},
  {Belloche}, {Hennebelle}, {Cabrit}, {Codella}, {Gueth}, {Podio}, {Anderl},
  {Bacmann}, {Bontemps}, {Gaudel}, {Ladjelate}, {Lef{\`e}vre}, {Tabone}, \&
  {Lefloch}}]{Maury:2019}
{Maury}, A.~J., {Andr{\'e}}, P., {Testi}, L., {et~al.} 2019, \aap, 621, A76

\bibitem[{{Minissale} {et~al.}(2016){Minissale}, {Dulieu}, {Cazaux}, \&
  {Hocuk}}]{Minissale:2016}
{Minissale}, M., {Dulieu}, F., {Cazaux}, S., \& {Hocuk}, S. 2016, \aap, 585,
  A24

\bibitem[{{Mizuno} {et~al.}(1994){Mizuno}, {Onishi}, {Hayashi}, {Ohashi},
  {Sunada}, {Hasegawa}, \& {Fukui}}]{mizuno1994}
{Mizuno}, A., {Onishi}, T., {Hayashi}, M., {et~al.} 1994, \nat, 368, 719

\bibitem[{{Murillo} \& {Lai}(2013)}]{Murillo:2013}
{Murillo}, N.~M. \& {Lai}, S.-P. 2013, \apjl, 764, L15

\bibitem[{{Oba} {et~al.}(2018){Oba}, {Tomaru}, {Lamberts}, {Kouchi}, \&
  {Watanabe}}]{Oba:2018}
{Oba}, Y., {Tomaru}, T., {Lamberts}, T., {Kouchi}, A., \& {Watanabe}, N. 2018,
  Nature Astronomy, 2, 228

\bibitem[{{Ohashi} {et~al.}(2018){Ohashi}, {Sanhueza}, {Sakai}, {Kand ori},
  {Choi}, {Hirota}, {Nguyễn-Lu'o'ng}, \& {Tatematsu}}]{Ohashi:2018}
{Ohashi}, S., {Sanhueza}, P., {Sakai}, N., {et~al.} 2018, \apj, 856, 147

\bibitem[{{Omukai}(2007)}]{Omukai:2007}
{Omukai}, K. 2007, \pasj, 59, 589

\bibitem[{{Onishi} {et~al.}(1999){Onishi}, {Mizuno}, {Kawamura}, \&
  {Fukui}}]{onishi1999}
{Onishi}, T., {Mizuno}, A., {Kawamura}, A., \& {Fukui}, Y. 1999, in Star
  Formation 1999, ed. T.~{Nakamoto}, 153--158

\bibitem[{{Onishi} {et~al.}(1996){Onishi}, {Mizuno}, {Kawamura}, {Ogawa}, \&
  {Fukui}}]{onishi1996}
{Onishi}, T., {Mizuno}, A., {Kawamura}, A., {Ogawa}, H., \& {Fukui}, Y. 1996,
  \apj, 465, 815

\bibitem[{{Onishi} {et~al.}(1998){Onishi}, {Mizuno}, {Kawamura}, {Ogawa}, \&
  {Fukui}}]{onishi1998}
{Onishi}, T., {Mizuno}, A., {Kawamura}, A., {Ogawa}, H., \& {Fukui}, Y. 1998,
  \apj, 502, 296

\bibitem[{{Onishi} {et~al.}(2002){Onishi}, {Mizuno}, {Kawamura}, {Tachihara},
  \& {Fukui}}]{onishi2002}
{Onishi}, T., {Mizuno}, A., {Kawamura}, A., {Tachihara}, K., \& {Fukui}, Y.
  2002, \apj, 575, 950

\bibitem[{{Punanova} {et~al.}(2018){Punanova}, {Caselli}, {Feng},
  {Chac{\'o}n-Tanarro}, {Ceccarelli}, {Neri}, {Fontani}, {Jim{\'e}nez-Serra},
  {Vastel}, {Bizzocchi}, {Pon}, {Vasyunin}, {Spezzano}, {Hily-Blant}, {Testi},
  {Viti}, {Yamamoto}, {Alves}, {Bachiller}, {Balucani}, {Bianchi},
  {Bottinelli}, {Caux}, {Choudhury}, {Codella}, {Dulieu}, {Favre}, {Holdship},
  {Jaber Al-Edhari}, {Kahane}, {Laas}, {LeFloch}, {L{\'o}pez-Sepulcre},
  {Ospina-Zamudio}, {Oya}, {Pineda}, {Podio}, {Quenard}, {Rimola}, {Sakai},
  {Sims}, {Taquet}, {Theul{\'e}}, \& {Ugliengo}}]{Punanova:2018}
{Punanova}, A., {Caselli}, P., {Feng}, S., {et~al.} 2018, \apj, 855, 112

\bibitem[{{Qu{\'e}nard} {et~al.}(2017){Qu{\'e}nard}, {Vastel}, {Ceccarelli},
  {Hily-Blant}, {Lefloch}, \& {Bachiller}}]{quenard2017a}
{Qu{\'e}nard}, D., {Vastel}, C., {Ceccarelli}, C., {et~al.} 2017, \mnras, 470,
  3194

\bibitem[{{Rimola} {et~al.}(2014){Rimola}, {Taquet}, {Ugliengo}, {Balucani}, \&
  {Ceccarelli}}]{Rimola:2014}
{Rimola}, A., {Taquet}, V., {Ugliengo}, P., {Balucani}, N., \& {Ceccarelli}, C.
  2014, \aap, 572, A70

\bibitem[{{Schnee} {et~al.}(2012){Schnee}, {Sadavoy}, {Di Francesco},
  {Johnstone}, \& {Wei}}]{Schnee:2012}
{Schnee}, S., {Sadavoy}, S., {Di Francesco}, J., {Johnstone}, D., \& {Wei}, L.
  2012, \apj, 755, 178

\bibitem[{{Shinnaga} {et~al.}(2009){Shinnaga}, {Phillips}, {Furuya}, \&
  {Kitamura}}]{shinnaga2009}
{Shinnaga}, H., {Phillips}, T.~G., {Furuya}, R.~S., \& {Kitamura}, Y. 2009,
  \apjl, 706, L226

\bibitem[{{Soma} {et~al.}(2015){Soma}, {Sakai}, {Watanabe}, \&
  {Yamamoto}}]{Soma:2015}
{Soma}, T., {Sakai}, N., {Watanabe}, Y., \& {Yamamoto}, S. 2015, \apj, 802, 74

\bibitem[{{Tafalla} {et~al.}(2004){Tafalla}, {Myers}, {Caselli}, \&
  {Walmsley}}]{tafalla2004}
{Tafalla}, M., {Myers}, P.~C., {Caselli}, P., \& {Walmsley}, C.~M. 2004, \apss,
  292, 347

\bibitem[{{Tafalla} {et~al.}(2002){Tafalla}, {Myers}, {Caselli}, {Walmsley}, \&
  {Comito}}]{Tafalla:2002}
{Tafalla}, M., {Myers}, P.~C., {Caselli}, P., {Walmsley}, C.~M., \& {Comito},
  C. 2002, \apj, 569, 815

\bibitem[{{Takahashi} {et~al.}(2013){Takahashi}, {Ohashi}, \&
  {Bourke}}]{takahashi2013}
{Takahashi}, S., {Ohashi}, N., \& {Bourke}, T.~L. 2013, \apj, 774, 20

\bibitem[{{Tokuda} {et~al.}(2016){Tokuda}, {Onishi}, {Matsumoto}, {Saigo},
  {Kawamura}, {Fukui}, {Inutsuka}, {Machida}, {Tomida}, {Tachihara}, \&
  {Andr{\'e}}}]{Tokuda:2016}
{Tokuda}, K., {Onishi}, T., {Matsumoto}, T., {et~al.} 2016, \apj, 826, 26

\bibitem[{{Tokuda} {et~al.}(2017){Tokuda}, {Onishi}, {Saigo}, {Hosokawa},
  {Matsumoto}, {Inutsuka}, {Machida}, {Tomida}, {Kunitomo}, {Kawamura},
  {Fukui}, \& {Tachihara}}]{Tokuda:2017}
{Tokuda}, K., {Onishi}, T., {Saigo}, K., {et~al.} 2017, \apj, 849, 101

\bibitem[{{Tokuda} {et~al.}(2014){Tokuda}, {Onishi}, {Saigo}, {Kawamura},
  {Fukui}, {Matsumoto}, {Inutsuka}, {Machida}, {Tomida}, \&
  {Tachihara}}]{tokuda2014}
{Tokuda}, K., {Onishi}, T., {Saigo}, K., {et~al.} 2014, \apjl, 789, L4

\bibitem[{{Tokuda} {et~al.}(2018){Tokuda}, {Onishi}, {Saigo}, {Matsumoto},
  {Inoue}, {Inutsuka}, {Fukui}, {Machida}, {Tomida}, {Hosokawa}, {Kawamura}, \&
  {Tachihara}}]{Tokuda:2018}
{Tokuda}, K., {Onishi}, T., {Saigo}, K., {et~al.} 2018, \apj, 862, 8

\bibitem[{{Tomida} {et~al.}(2010){Tomida}, {Machida}, {Saigo}, {Tomisaka}, \&
  {Matsumoto}}]{Tomida:2010}
{Tomida}, K., {Machida}, M.~N., {Saigo}, K., {Tomisaka}, K., \& {Matsumoto}, T.
  2010, \apjl, 725, L239

\bibitem[{{Turner} \& {Apponi}(2001)}]{Turner:2001}
{Turner}, B.~E. \& {Apponi}, A.~J. 2001, \apjl, 561, L207

\bibitem[{{van der Tak} {et~al.}(2007){van der Tak}, {Black}, {Sch{\"o}ier},
  {Jansen}, \& {van Dishoeck}}]{van-der-Tak:2007}
{van der Tak}, F.~F.~S., {Black}, J.~H., {Sch{\"o}ier}, F.~L., {Jansen}, D.~J.,
  \& {van Dishoeck}, E.~F. 2007, \aap, 468, 627

\bibitem[{{Vastel} {et~al.}(2014){Vastel}, {Ceccarelli}, {Lefloch}, \&
  {Bachiller}}]{Vastel:2014}
{Vastel}, C., {Ceccarelli}, C., {Lefloch}, B., \& {Bachiller}, R. 2014, \apjl,
  795, L2

\bibitem[{{Vastel} {et~al.}(2016){Vastel}, {Ceccarelli}, {Lefloch}, \&
  {Bachiller}}]{vastel2016}
{Vastel}, C., {Ceccarelli}, C., {Lefloch}, B., \& {Bachiller}, R. 2016, \aap,
  591, L2

\bibitem[{{Vastel} {et~al.}(2019){Vastel}, {Loison}, {Wakelam}, \&
  {Lefloch}}]{vastel2019}
{Vastel}, C., {Loison}, J.~C., {Wakelam}, V., \& {Lefloch}, B. 2019, \aap, 625,
  A91

\bibitem[{{Vastel} {et~al.}(2018){Vastel}, {Qu{\'e}nard}, {Le Gal}, {Wakelam},
  {Andrianasolo}, {Caselli}, {Vidal}, {Ceccarelli}, {Lefloch}, \&
  {Bachiller}}]{vastel2018b}
{Vastel}, C., {Qu{\'e}nard}, D., {Le Gal}, R., {et~al.} 2018, \mnras, 478, 5514

\bibitem[{{Vasyunin} {et~al.}(2017){Vasyunin}, {Caselli}, {Dulieu}, \&
  {Jim{\'e}nez-Serra}}]{vasyunin2017}
{Vasyunin}, A.~I., {Caselli}, P., {Dulieu}, F., \& {Jim{\'e}nez-Serra}, I.
  2017, \apj, 842, 33

\bibitem[{{Vidal} \& {Wakelam}(2018)}]{Wakelam:2018}
{Vidal}, T. H.~G. \& {Wakelam}, V. 2018, \mnras, 474, 5575

\bibitem[{{Vorobyov} {et~al.}(2017){Vorobyov}, {Elbakyan}, {Dunham}, \&
  {Guedel}}]{Vorobyov:2017}
{Vorobyov}, E.~I., {Elbakyan}, V., {Dunham}, M.~M., \& {Guedel}, M. 2017, \aap,
  600, A36

\bibitem[{{Wakelam} {et~al.}(2004){Wakelam}, {Caselli}, {Ceccarelli}, {Herbst},
  \& {Castets}}]{Wakelam:2004}
{Wakelam}, V., {Caselli}, P., {Ceccarelli}, C., {Herbst}, E., \& {Castets}, A.
  2004, \aap, 422, 159

\bibitem[{{Watanabe} \& {Kouchi}(2002)}]{Watanabe:2002}
{Watanabe}, N. \& {Kouchi}, A. 2002, \apjl, 571, L173

\bibitem[{{Young} {et~al.}(2004){Young}, {J{\o}rgensen}, {Shirley},
  {Kauffmann}, {Huard}, {Lai}, {Lee}, {Crapsi}, {Bourke}, {Dullemond},
  {Brooke}, {Porras}, {Spiesman}, {Allen}, {Blake}, {Evans}, {Harvey},
  {Koerner}, {Mundy}, {Myers}, {Padgett}, {Sargent}, {Stapelfeldt}, {van
  Dishoeck}, {Bertoldi}, {Chapman}, {Cieza}, {DeVries}, {Ridge}, \&
  {Wahhaj}}]{Young:2004}
{Young}, C.~H., {J{\o}rgensen}, J.~K., {Shirley}, Y.~L., {et~al.} 2004, \apjs,
  154, 396

\end{thebibliography}

%===============================================================
%===============================================================

\end{document}